\documentclass[12pt]{article}

\usepackage{scicite,epsfig,amssymb,amsmath,lscape}

\usepackage{times}
\usepackage{caption}

\usepackage{hyperref}

\newcommand{\begit}{\begin{itemize}}
\newcommand{\enit}{\end{itemize}}
\newcommand{\begen}{\begin{enumerate}}
\newcommand{\enen}{\end{enumerate}}

\newcommand{\beq}{\begin{equation}}
\newcommand{\eeq}{\end{equation}}
\newcommand{\beqa}{\begin{eqnarray}} 
\newcommand{\eeqa}{\end{eqnarray}}

\newdimen\sa  \newdimen\sb
\def\parcs{\sa=.07em \sb=.03em
     \ifmmode $\rlap{.}$^{\scriptscriptstyle\prime\kern -\sb\prime}$\kern -\sa$
     \else \rlap{.}$^{\scriptscriptstyle\prime\kern -\sb\prime}$\kern -\sa\fi}

\newenvironment{sciabstract}{%
\begin{quote} \bf}
{\end{quote}}

\newcounter{lastnote}

\title{Discovery of a Candidate Black Hole$-$Giant Star Binary System in the Galactic Field} 

\author
{Todd A.\ Thompson,$^{1,2,3\ast}$ 
Christopher S.\ Kochanek,$^{1,2}$
Krzysztof Z.\ Stanek,$^{1,2}$   \\
Carles\ Badenes,$^{4,5}$ 
Richard S.\ Post,$^{6}$ 
Tharindu Jayasinghe,$^{1,2}$ \\
David W.\ Latham,$^{7}$ Allyson Bieryla,$^{7}$ 
Gilbert A. Esquerdo,$^{7}$ \\
Perry Berlind,$^{7}$ Michael L. Calkins,$^{7}$
Jamie Tayar,$^{1}$
Lennart Lindegren,$^{8}$\\
Jennifer A.\ Johnson,$^{1,2}$
Thomas W.-S.\ Holoien,$^{9}$ 
Katie Auchettl,$^{2,10}$
Kevin Covey$^{11}$}

\date{}

\begin{document} 


\baselineskip24pt


\maketitle 

\noindent
\normalsize{$^{1}$ Department of Astronomy, The Ohio State University, 140 W.\ 18th Ave., Columbus, OH 43210, USA}\\
\normalsize{$^{2}$ Center for Cosmology and AstroParticle Physics (CCAPP), The Ohio State University, 191 W.\ Woodruff Ave., Columbus, OH 43210, USA } \\
\normalsize{$^{3}$ Institute for Advanced Study, 1 Einstein Drive, Princeton, NJ 08540, USA} \\
\normalsize{$^{4}$Department of Physics and Astronomy and Pittsburgh Particle Physics, Astrophysics and Cosmology Center (PITT PACC), University of Pittsburgh, 3941 O'Hara Street, Pittsburgh, PA 15260, USA}\\
\normalsize{$^{5}$  Institut de Ci\`encies del Cosmos (ICCUB), Universitat de Barcelona (IEEC-UB), Mart\'i Franqu\'es 1, E08028 Barcelona, Spain} \\
\normalsize{$^{6}$ Post Observatory, Lexington, MA 02421, USA}\\ 
\normalsize{$^{7}$ Harvard-Smithsonian Center for Astrophysics, Cambridge, MA 02138, USA}\\
\normalsize{$^{8}$ Lund Observatory, Department of Astronomy and Theoretical Physics, Lund University, Box 43, 22100 Lund, Sweden}\\
\normalsize{$^{9}$   The Observatories of the Carnegie Institution for Science, 813 Santa Barbara St., Pasadena, CA 91101, USA}\\
\normalsize{$^{10}$ Department of Physics, The Ohio State University, 191 W. Woodruff Avenue, Columbus, OH 43210, USA}  \\
\normalsize{$^{11}$ Department of Physics and Astronomy, Western Washington University, Bellingham, WA, 98225, USA} \\
\normalsize{$^\ast$To whom correspondence should be addressed; E-mail: thompson.1847@osu.edu}

\baselineskip24pt
\begin{sciabstract}
We report the discovery of the first likely black hole in a non-interacting binary system with a field red giant. By combining radial velocity measurements from the Apache Point Observatory Galactic Evolution Experiment (APOGEE) with photometric variability data from the All-Sky Automated Survey for Supernovae (ASAS-SN), we identified the bright rapidly-rotating giant 2MASS J05215658+4359220 as a binary system with a massive unseen companion. Subsequent radial velocity measurements reveal a system with an orbital period of $\sim83$\,days and near-zero eccentricity. The photometric variability period of the giant is consistent with the orbital period, indicative of star spots and tidal synchronization. Constraints on the giant's mass and radius from its luminosity, surface gravity, and temperature imply an unseen companion with mass of $3.3^{+2.8}_{-0.7}$\,M$_\odot$, indicating a low-mass black hole or an exceedingly massive neutron star. Measurement of the astrometric binary motion by {\it Gaia} will further characterize the system. This discovery demonstrates the potential of massive spectroscopic surveys like APOGEE and all-sky, high-cadence photometric surveys like ASAS-SN to revolutionize our understanding of the compact object mass function, and to test theories of binary star evolution and the supernova mechanism.
\end{sciabstract}

The neutron star and stellar black hole mass functions are the subject of intense study as they directly constrain the mechanism of core-collapse supernovae, its success and failure rate as a function of metallicity, and the physics of binary stars \cite{Pejcha_landscape,PejchaNS,Belczynski}.  An unbiased census of neutron star and black hole masses is critical to a broad swath of astrophysics.

To date, however, our knowledge of neutron star and black hole demography is limited. In particular, mass measurements come almost exclusively from pulsar and accreting binary systems selected from radio, X-ray, and gamma-ray surveys \cite{OzelNS,OzelBH}. While Galactic gravitational microlensing has the potential to reveal compact object mass distributions directly \cite{Paczynski,Wyrzykowski}, most systems cannot be followed up. The recent discovery of merging black hole and neutron star binaries by LIGO \cite{LIGO_BH,LIGO_NS} provides a new window on compact object masses, but these systems are an intrinsically biased subset of the parent population. 

Studies of compact object populations are complemented by those of massive star binary systems \cite{kobulnicky,Sana}, which indicate a broad distribution of secondary masses and orbital periods, implying that many massive stars should have low-mass companions. Yet, quiescent black hole stellar binaries born in the field have not been discovered in directed radial velocity searches, even though the promise of such systems has been discussed for decades \cite{Guseinov,Trimble}. The one serendipitous discovery to date likely formed by dynamical processes in a globular cluster \cite{Giesers}. Although subject to their own selection biases, a large collection of Galactic binary star systems with black hole or neutron star companions would provide a wholly new population for study and for comparison with binary stellar evolution models. Because field binaries are coeval, such systems would constrain the black hole production rate as a function of metallicity and age, and thus give vital clues to critically uncertain aspects of binary evolution.

To address these issues, we initiated a search for binary systems with massive unseen companions in data from APOGEE \cite{Majewski,Zasowski,Abolfathi}, part of the Sloan Digital Sky Survey IV \cite{Gunn,Blanton}. APOGEE provides high signal-to-noise, multi-epoch near-infrared spectroscopy for over $10^5$ stars throughout the Galaxy. These observations provide an unprecedented accounting of the chemical abundances of stars and can also reveal radial velocity variations indicative of binary orbital motion. Using the catalog constructed by \cite{Badenes}, we searched for systems with large apparent accelerations --- the difference in radial velocity divided by the difference in time between two epochs. Systems were then rank ordered by an estimate of the binary mass function given the uneven timing of the APOGEE epochs. 

Although the radial velocity measurements from APOGEE can immediately indicate the presence of a binary, the mass of the companion is uncertain because the orbital period, inclination, and eccentricity are unknown. To constrain the orbital periods of the $\sim200$ APOGEE sources with the highest accelerations, we searched for periodic photometric variations in data from the All-Sky Automated Survey for Supernovae (ASAS-SN) \cite{Shappee14,Kochanek17} that might be indicative of transits, ellipsoidal variations, or starspots. This process immediately yielded the giant 2MASS J05215658+4359220 (hereafter ``J05215658"), which exhibits an acceleration of $\simeq2.9$\,km/s/day and periodic photometric variability with a period of $\sim82.2\pm2.5$\,days. The system lies towards Auriga, near the outer Galactic plane, with Galactic coordinates $(l,b)=(164.774^\circ,  4.184^\circ)$. With a mean visual magnitude of $V\simeq12.9$, the system appears in many archival catalogs. 

The combined APOGEE spectrum of J05215658 gives $T_{\rm eff}\simeq4480\pm62$\,K, $\log g \simeq 2.59\pm0.06$, $[{\rm M/H}]\simeq-0.30\pm0.03$, $[{\rm \alpha/M}]\simeq -0.04\pm0.015$ and $[{\rm C/N}]\simeq0.0$. Our analysis of the APOGEE spectra yields a projected rotation velocity of $v\sin i_{\rm rot}\simeq14.1\pm0.6$\,km s$^{-1}$ \cite{Tayar}. The optical spectra we report below from the Tillinghast Reflector Echelle Spectrograph (TRES) give values of the temperature and metallicity consistent with APOGEE, but lower $\log g\simeq2.35\pm0.14$, and somewhat different $v\sin i_{\rm rot}\simeq16.8\pm0.6$\,km s$^{-1}$. Because APOGEE systematically overestimates $\log g$ for stars with large $v\sin i_{\rm rot}$ and because it has lower spectral resolution, we use the TRES $\log g$ in our analysis. Because the TRES $v\sin i_{\rm rot}$ does not include a correction for macroturbulent broadening, we adopt the APOGEE $v\sin i_{\rm rot}$. 

Assuming that the giant has a mass $M_{\rm giant}\ge1$\,M$_\odot$, and that the $\sim82$\,day photometric variability is either ellipsoidal variations or starspots in a tidally synchronized binary, the APOGEE radial velocities alone imply a minimum mass for the unseen companion above the Chandrasekhar mass of $\simeq1.4$\,M$_\odot$, ruling out a white dwarf. 

To further constrain the system, we initiated radial velocity and multi-band photometric followup, shown in Figure \ref{figure:post}. The photometric variability amplitude increases from the redder (top) to bluer (bottom) bands, and is phased with the orbital motion of the binary such that the maximum blueshift closely aligns with the photometric maximum in all bands. The shape, character, and phasing of the lightcurve are inconsistent with stellar pulsations or ellipsoidal variations, but are typical of the class of spotted K giants like HD 1833, V1192 Orionis, and KU Pegasi \cite{Strassmeier,Weber_Strassmeier}. Such systems in binaries with periods less than $\sim150$\,days often have low eccentricities, implying rapid tidal circularization \cite{Mayor,Verbunt_Phinney,Strassmeier2012}. The change in the shape of the lightcurve over time shown in Figure \ref{figure:lc} likely indicates spot evolution.

The TRES radial velocity measurements are shown in the lower panel of Figure~\ref{figure:post},. Like the APOGEE spectra, the TRES spectra always exhibit only a single set of absorption lines, indicative of a single-lined spectroscopic binary. The system has a nearly circular orbit with $P_{\rm orb}\simeq83.2\pm0.06$\,days, radial velocity semi-amplitude $K\simeq44.6\pm0.1$\,km s$^{-1}$, and eccentricity $e\simeq0.0048\pm0.0026$. The mass function is 
\beq
f(M)=\frac{M_{\rm CO}^3\sin^3 i_{\rm orb}}{(M_{\rm giant}+M_{\rm CO})^2}=\frac{K^3P_{\rm orb}}{2\pi G}(1-e^2)^{3/2}\simeq0.77\,M_{\rm \odot},
\label{mass_function}
\eeq
where $M_{\rm CO}$ is the mass of the compact object companion and $i_{\rm orb}$ is the orbital inclination. Solutions to $f(M)$ for $M_{\rm CO}$ as a function of $M_{\rm giant}$ and several values of $\sin i_{\rm orb}$ are shown in Figure \ref{figure:m} (solid black lines). For $M_{\rm giant}\ge1$\,M$_\odot$ and $\sin i_{\rm orb}=1.0$, the minimum possible companion mass is $M_{\rm CO}\ge1.8$\,M$_\odot$. The observed spectral energy distribution (SED) is inconsistent with a stellar companion of such high mass.  We conclude that the unseen companion is either a massive neutron star or a black hole.

In order to understand the nature of the system, we must constrain the mass of the giant and the orbital inclination. We provide several arguments that suggest the giant is an intermediate mass star with $M_{\rm giant}\sim2-4$\,M$_\odot$. Figure \ref{figure:m} then implies either an unprecedentedly massive neutron star or a low-mass black hole.

We first estimate the giant radius and luminosity using $v\sin i$. The low orbital eccentricity and the correspondence between the orbital and photometric periods imply that the system is tidally circularized and synchronized. We thus adopt the hypothesis that the giant's rotational period is equal to the binary orbital period and that their inclinations on the sky are identical. Combining the giant's  $v \sin i\simeq14.1\pm0.6$\,km s$^{-1}$ with the period yields a minimum stellar radius of $R\simeq23\pm1\,R_\odot/\sin i$, a minimum bolometric luminosity $L=4\pi R^2\sigma_{\rm SB} T_{\rm eff}^4\simeq210\pm20\,L_\odot/\sin^2 i$ for $T_{\rm eff}\simeq4500$\,K, and a minimum distance $D\simeq [L/(4\pi F)]^{1/2}\simeq2.45\pm0.1\,{\rm kpc}/\sin i$, where $\sigma_{\rm SB}$ is the Stefan-Boltzmann constant and $F$ is the measured bolometric flux. Combining the minimum giant radius with the TRES $\log g\simeq2.35\pm0.14$ gives an estimate for the giant mass of $M_{\rm giant}^{\log g}=gR^2/G\simeq4.4_{-1.5}^{+2.2}$\,M$_\odot/\sin^2 i$, implying a minimum companion mass of $M_{\rm CO}\gtrsim2.9$\,M$_\odot$. 

The radius of the giant can also be determined directly from its measured distance and bolometric flux. The nominal Gaia parallax is $0.272\pm0.049$\,mas \cite{Lindegren} (ID 207628584632757632), but there are systematic offsets in the Gaia parallaxes that are a function of both sky position and the apparent magnitude. Additionally, the giant will exhibit astrometric motion that is not accounted for in Gaia's Data Release 2. Given $f(M)$ and $P$, the ratio of the binary angular motion to the parallax is $0.34/\sin i$, which could bias the measured parallax. Applying a zero-point offset and an additional systematic uncertainty on the basis of other studies, and accounting for the phased binary motion with the geometry and timing of the observations by Gaia for J05215658 using Monte Carlo simulations for an arbitrary orbital sky projection, we find a parallax of $\pi\simeq0.322^{+0.086}_{-0.074}$\,mas (1-$\sigma$ confidence interval) for all $\sin i>0$, corresponding to a distance of $D\simeq3.11^{+0.93}_{-0.66}$\,kpc, which, when combined with the bolometric flux gives $L\simeq331_{-127}^{+231}$\,L$_\odot$ and $R\simeq30^{+9}_{-6}$\,R$_\odot$, notably consistent with the corresponding numbers derived from $v \sin i$. Combining the Gaia lower bound on $R$ with the TRES $\log g$ gives $M_{\rm giant}^{\log g}\gtrsim4.8$\,M$_\odot$ and a value of $M_{\rm CO}$ in the black hole regime (Fig.\ \ref{figure:m}).  Direct comparison between $R\simeq30^{+9}_{-6}$ derived from Gaia and $R\simeq23\pm1\,R_\odot/\sin i$ derived from $v\sin i$ suggests that $\sin i \simeq 0.8\pm0.2$. 

Because of the low noise in the Gaia DR2 single-star astrometric solution, large biases in the observed parallax can be ruled out. We derive a 2-$\sigma$ upper limit on the Gaia parallax of $0.486$\,mas for $\sin i>0.6$, corresponding to 2-$\sigma$ lower limits of $L\gtrsim150$\,L$_\odot$ and $R\gtrsim20$\,R$_\odot$. The TRES $\log g=2.35\pm0.14$ then gives $M_{\rm giant}^{\log g}\gtrsim3.2_{-0.9}^{+1.2}$\,M$_\odot$, implying a lower limit on the companion mass of $M_{\rm CO}\gtrsim2.5$\,M$_\odot$ (Fig.\ \ref{figure:m}).

The giant mass can also be estimated by comparing its properties to single-star evolutionary tracks, with the caveats that (1) strong binary interaction likely occurred in the history of the system \cite{Linares}, and (2) that rapidly rotating, spotted giants like J05215658 are observed to be redder than expected for a given luminosity, implying a lower mass than the true dynamical mass as determined from eclipsing binary systems \cite{olah}. Nevertheless, we searched for the best fitting PARSEC \cite{Marigo} and MIST \cite{Choi} evolutionary tracks with metallicity $[{\rm Fe/H}]=0.0$, $-0.4$, and $-0.8$ and with the constraint $\log g=2.35\pm0.14$, and the other parameters inferred from Gaia and the SED ($L$, $R$, and $T_{\rm eff}$). The primary tension in matching the data is the difficulty of simultaneously fitting the temperature and gravity, perhaps a consequence of the known systematic differences between observed and modeled giant temperatures \cite{olah,Tayar}. The best matches were for Solar metallicity models, and worsen at lower metallicity. Combining all of the PARSEC and MIST models, we find a best joint fit of $M_{\rm giant}\simeq3.2^{+1.0}_{-1.0}$\,M$_\odot$ (2-$\sigma$), labelled by the vertical blue band labeled ``L, T$_{\rm eff}$" in Figure \ref{figure:m}. In general, the fits are driven upwards in giant mass by the $L$ and $R$ inferred from Gaia and the SED. Combining the fit value for $R$ with the $v\sin i$ derived from APOGEE gives a constraint on $\sin i\simeq0.97^{+0.03}_{-0.12}$. Solving the mass function $f(M)$ for this range of $M_{\rm giant}$ and $\sin i$ yields an unseen companion compact object mass of $M_{\rm CO}\simeq3.3^{+2.8}_{-0.7}$\,M$_\odot$  (2-$\sigma$). These bounds and the best fit are shown as the dotted black boxes with central black circle in Figure \ref{figure:m}. All models in the indicated parameter regime should exhibit $\sim1-2$\% ellipsoidal variability, which we cannot at present convincingly separate from overtones of the higher amplitude spot modulation in the ASAS-SN photometric data. 

These results are qualitatively insensitive to $v \sin i$. For the fits described, lower mass giants than the $3.2^{+1.0}_{-1.0}$\,M$_\odot$ indicated in Figure \ref{figure:m} are excluded because the inclination inferred from the giant radius and $v \sin i$ becomes unphysical ($\sin i > 1$).  A lower assumed $v \sin i$ (e.g., if macroturbulent line broadening is especially large in J05215658), allows for somewhat lower mass giants, but does not lead to solutions with lower mass companions: while lower $M_{\rm giant}$ does push $M_{\rm CO}$ lower, this is more than compensated for by the smaller inferred value of $\sin i$ that drives $M_{\rm CO}$ upwards. For example, allowing for $v \sin i$ as low as $10$\,km s$^{-1}$, our fits combining the evolutionary tracks with the Gaia parallax can give giant masses of $\simeq1.8$\,M$_\odot$, but never yield a best-fit value of $M_{\rm CO}$ below 2.5\,M$_\odot$. Similarly, our results from fitting to evolutionary tracks are robust to changes in $\log g$. If instead of using the TRES $\log g$, we instead impose $\log g=2.0\pm0.2$, or impose no constraint on $\log g$, we obtain best-fit giant masses at the low end of the range denoted in Figure \ref{figure:m} ($M_{\rm giant}\simeq2.2-2.5$\,M$_\odot$), but with compact object companion masses of $\simeq2.9-4.0$\,M$_\odot$.

Given the radii and luminosities inferred from both $v \sin i$ and Gaia, the stellar models require giant masses of $\sim 2-4$\,M$_\odot$ for the giant.  Although these stars are rare in APOGEE, they do exist in the sample. There are $135$ stars with astroseismic masses above $2.5$\,M$_\odot$ in the $\simeq6700$ object Apache Point Observatory-{\it Kepler} Asteroseismology Science Consortium (APOKASC) database \cite{Pinsonneault18}. However, J05215658 would be an outlier even among APOGEE's massive giants due to its high measured $[{\rm C/N}]$ ratio of $\simeq0.0$. Giants exhibit a negative correlation between $[{\rm C/N}]$ and mass as a result of nuclear processing in their interiors. Thus, even considering the uncertainties in the $[{\rm C/N}]$ determination (see Supplementary Material), a {\it prima facie} comparison of the mean trend from APOKASC with J05215658 implies a low value of $M_{\rm giant}\simeq1.0$\,M$_\odot$ and a corresponding minimum compact object mass of $M_{\rm CO}\simeq1.8$\,M$_\odot$ ($\sin i=1$). We highlight this ``low-mass giant" possibility with the vertical band labelled ``$[{\rm C/N}]$" for completeness in Figure \ref{figure:m}, but view it as very unlikely. The combination of the well-measured $T_{\rm eff}$ and the lower limits on the luminosity from Gaia effectively rule out low mass $\simeq1$\,M$_\odot$ solutions for the giant when compared to the rest of the APOKASC sample, or to standard evolutionary tracks. 

Instead, J05215658 is most likely simply an outlier from the mean $[{\rm C/N}]-M_{\rm giant}$ locus, perhaps owing to binary interactions. A handful of stars in the APOKASC sample have astroseismic $M_{\rm giant}> 2.0$\,M$_\odot$ {\it and} $[{\rm C/N}]\ge-0.1$, and the fraction of such stars, while small, appears to grow with mass. This high-mass high-$[{\rm C/N}]$ sequence may arise from mergers of lower mass stars \cite{Izzard}. Perhaps a merger origin for the giant in J05215658  could explain its simultaneously high $L$ and high $[{\rm C/N}]$, at the cost of invoking a former triple system. More plausibly, the peculiar abundances of the giant may be the result of previous interaction with its binary companion, either via mass transfer or during the explosive event that may have accompanied the formation of the compact object. 

On the basis of all the current evidence, we conclude that the remarkable system J05215658 likely consists of a $\simeq3.2^{+1.0}_{-1.0}$\,M$_\odot$ giant and a low-mass black hole of $M_{\rm CO}\simeq3.3^{+2.8}_{-0.7}$\,M$_\odot$  (the 2-$\sigma$ range in Fig.\ \ref{figure:m}), encompassing a theoretical lower mass limit for black holes of $\sim4$\,M$_\odot$ obtained by some recent studies, and potentially below the lowest well-measured black hole mass to date \cite{OzelBH}. The lower mass range approaches the theoretical maximum neutron star mass of $\simeq2.5$\,M$_\odot$ \cite{Lattimer}, and would be higher than the maximum neutron star mass yet observed $\simeq2.0$\,M$_\odot$ \cite{Antoniadis} (see Fig.~\ref{figure:m}; but, see Ref.\ \cite{Breivik}). A further possibility is that J05215658 is a related evolutionary descendant of the recently discovered class of high-mass millisecond pulsar systems with distant main sequence binary companions like PSR J1903+0327 \cite{Champion}. Important additional constraints will be provided by Gaia through detection of astrometric motion, by TESS, which may provide an astroseismic constraint on the giant mass, and by improved photometry, which may convincingly reveal ellipsoidal variations.  

The implications of our findings are far-reaching. First, we demonstrate the key importance of combining massive multi-epoch spectroscopic surveys like APOGEE or the upcoming SDSS V with all-sky high-cadence imaging surveys like ASAS-SN. Combining data sets allows us to quickly isolate systems for followup using well-defined selection criteria. Second, we have discovered the first quiescent non-interacting neutron star or, more likely, black hole stellar binary system whose mass and orbit will test supernova and binary stellar evolution theories. In particular, the compact object we identify may be one of the most massive neutron stars or one of the lowest mass black holes ever found  \cite{OzelNS,OzelBH}. Third, our work illustrates the utility of bright giants for finding compact objects. With their rapid tidal circularization timescales and large physical sizes, they pick out new regimes of binary evolution and reveal their orbital period with their brightness changes, whether via spots or ellipsoidal variations, before initiating additional spectroscopic studies. Indeed, it may be efficient to select such objects for radial velocity followup from already completed photometric surveys on the basis of their photometric variability alone. In this way, like pulsars, spotted tidally-synchronized giants like J05215658, with their easily detected periodic photometric variations, may reveal a new compact object demography.

\bibliographystyle{Science}
\bibliography{auriga.bib}

\section*{Acknowledgement}
This work was supported in part by Scialog Scholar grant 24216 from the Research Corporation.  T.A.T. acknowledges partial support from a Simons Foundation Fellowship and an IBM Einstein Fellowship from the Institute for Advanced Study, Princeton. T.C.B. acknowledges partial support from PHY 14-30152: Physics Frontier Center/JINA Center for the Evolution of the Elements (JINA-CEE), awarded by the US NSF.  We thank Jieun Choi for making MIST models available and C.\ Jordi and C.\ Fabricius for discussions. We thank the Ohio State University College of Arts and Sciences Technology Services for setting up the ASAS-SN {\it Sky Patrol} lightcurve server, which was critically useful during this work.  T.A.T. thanks J.\ Zinn, T.\ Sukhbold, S.\ Gaudi, O.\ Pejcha, K.\ Stassun, M.\ Pinsonneault, A.\ Brown, C.\ Gammie, E.\ Rossi, J.\ Fuller, S.\ Phinney, A.-C.\ Eilers, D.\ Hogg, K.\ Cunha, and R.\ Poleski for discussions, and K.\ A.\ Byram for encouragement and support.

\clearpage

\begin{figure*}
\centerline{\includegraphics[width=10.75cm]{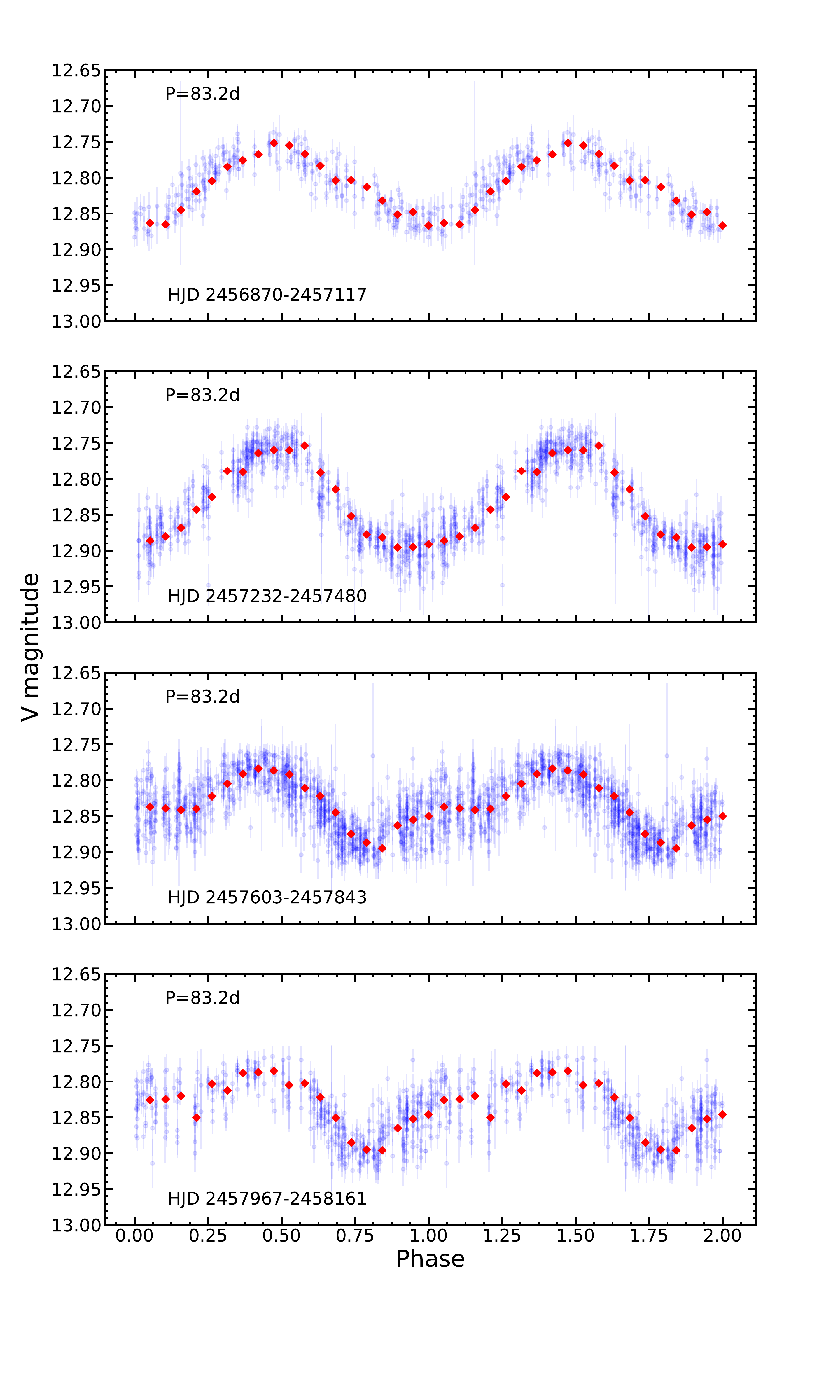}}
\vspace*{-1cm}
\caption{Multi-epoch V band ASAS-SN lightcurves over four observing seasons, phased to the orbital period of $83.2$\,days. Blue points are the data with error bars. Red points are a running median using 20 data points.}
\label{figure:lc}
\end{figure*}

\begin{figure*}
\centerline{\includegraphics[width=15.75cm]{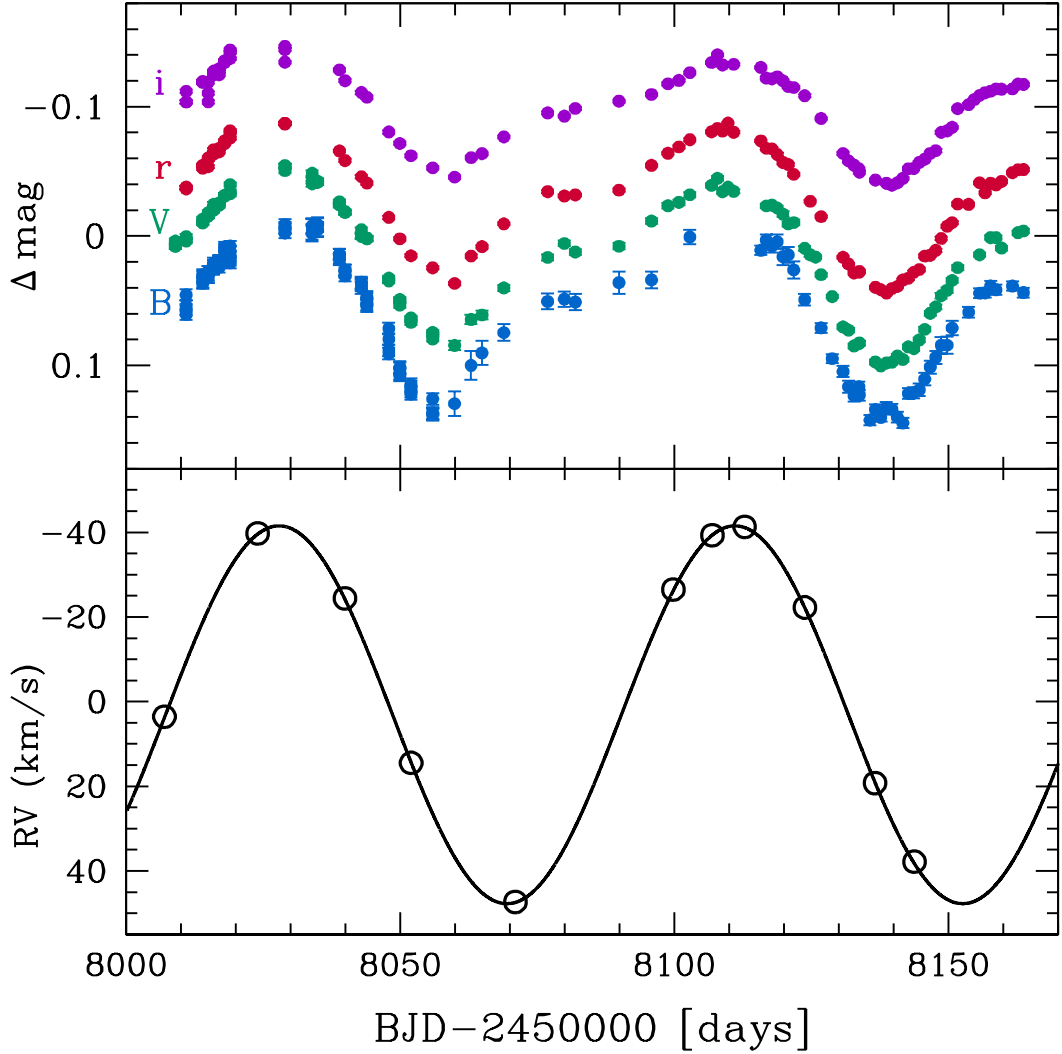}}
\vspace*{-3cm}
\caption{Multi-color $B$ (lowest), $V$, $r$, and $i$ (highest) lightcurves (top panel) with arbitrary zero-point offsets for clarity and the TRES radial velocity measurements (bottom panel). The phasing is such that maximum blueshift (negative RV) occurs very near the photometric maximum in all bands, and maximum redshift occurs after photometric minimum, near the ``shoulder/plateau" in the lightcurve at ${\rm BJD}-2450000\simeq8080$. Figure \ref{figure:lc} shows the evolution of the phased multi-epoch ASAS-SN lightcurve for comparison.}
\label{figure:post}
\end{figure*}
  
\begin{figure*}
\centerline{\includegraphics[width=14.5cm]{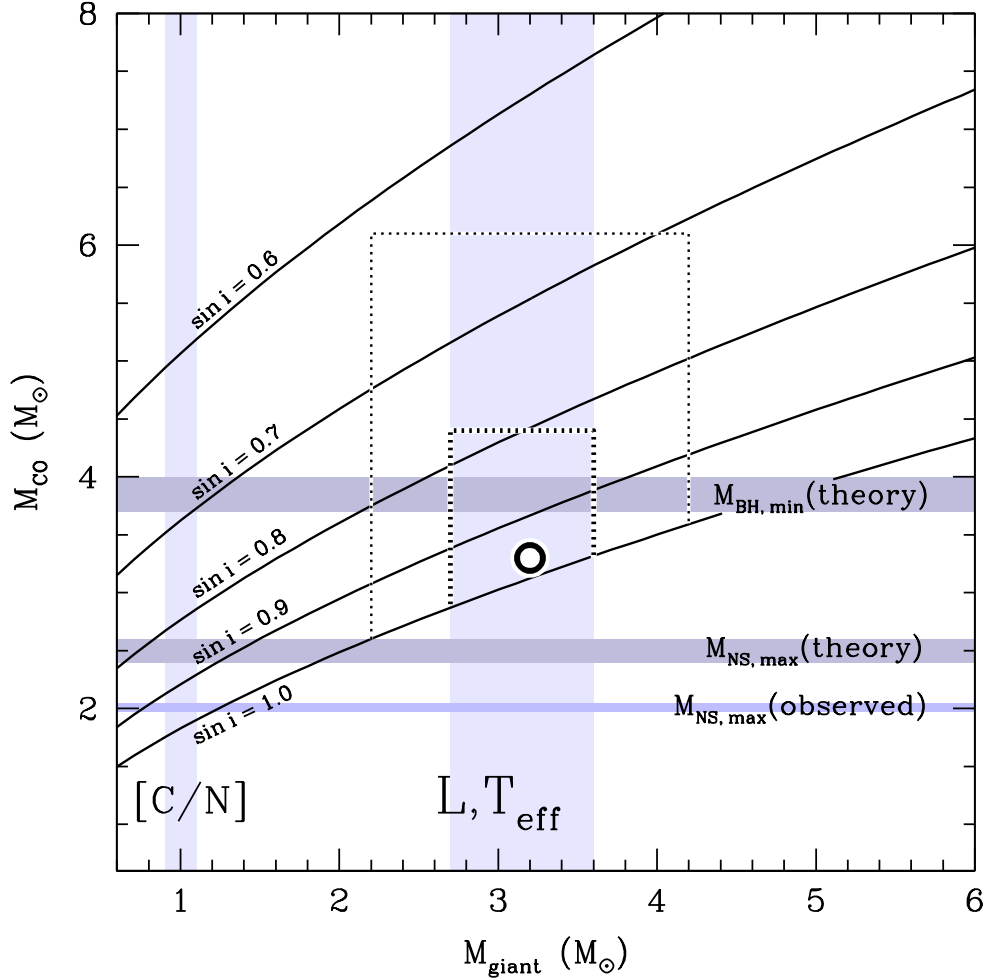}}
\vspace*{-3.75cm}
\caption{Solutions to the mass function for the compact object's mass $M_{\rm CO}$ as a function of the giant's mass $M_{\rm giant}$ with orbital inclinations of $\sin i=1.0-0.6$ (lowest to highest), shown as the solid black lines. The vertical band labeled ``L, T$_{\rm eff}$" denotes the best-fit range of $M_{\rm giant}$ when the giant's measured $L$, $T_{\rm eff}$, and $\log g$ are matched to theoretical single-star evolutionary tracks. The thick and thin dashed boxes shows the 1-$\sigma$ and 2-$\sigma$  mass ranges, respectively. The best fit is denoted with the black circle. The vertical band labeled  ``$[{\rm C/N}]$" denotes the range of $M_{\rm giant}$ using the mean locus of the observed $[{\rm C/N}]-M_{\rm giant}$ correlation for giant stars \cite{Pinsonneault18}. The horizontal bands denote the maximum neutron star mass so far observed (lowest; \cite{Antoniadis}, but see Ref.\ \cite{Breivik}), the theoretical maximum neutron star mass (middle; \cite{Lattimer}), and the minimum black hole mass from some recent theoretical models (top; \cite{Kochanek,Pejcha_landscape,Sukhbold}, but see \cite{Woosley1995,Kushnir}).}
\label{figure:m}
\end{figure*}
 
\clearpage
\setcounter{page}{1}
\begin{center}
\title{\bf {\Large Supplementary Materials for}\\[0.5cm]
 Discovery of an Extraordinary Binary System}

\author
{Thompson et al.} \\

\vspace*{0.5cm}

\normalsize{Correspondence to: thompson.1847@osu.edu.}
\newline
\end{center}

{{\bf This PDF file includes:}\\
\indent Supplementary text:\\
\indent \indent Object Selection Method\\
\indent \indent Multi-Band Optical Photometric Followup\\
\indent \indent Gaia Parallax, Proper Motion, and Binary Motion \\
\indent \indent Radial Velocity Followup\\
\indent \indent Properties of the Giant\\
\indent \indent Further Acknowledgement \\
\indent Figures:  \ref{figure:bias},  \ref{figure:ruwe},  \ref{figure:trueplx}, \ref{figure:vsini},  \ref{figure:rot}, \ref{figure:mist}, \ref{figure:sed}, \ref{figure:cn} \\
\indent Tables:  \ref{table:parallax}, \ref{table:rv}, \ref{table:orbital}, \ref{table:xirt}, \ref{table:xirt_apogee}, \ref{table:abundances}, \ref{table:uvm2}, \ref{table:giant_photometry}\\

\clearpage

\section{Object selection method}
\label{section:method}

The APOGEE survey provides multi-epoch spectroscopy for over $\sim10^5$ stars in the Galaxy. A catalog of high signal-to-noise radial velocity (RV) measurements was assembled by \cite{Badenes}. In general, there are $\sim2-4$ measurements per system. Although a measured RV difference between subsequent epochs can indicate the presence of a binary companion, the orbit is in general not well-established with such a small number of RV samplings \cite{PriceWhelan2017,PriceWhelan2018}. A simple criterion is used to identify systems that might have a massive unseen companion. We first calculated the maximum acceleration for each system,
\beq
{a}_{\rm max} = \left. \frac{\Delta{\rm RV}}{\Delta t_{\rm RV}}\right|_{\rm max}
\eeq
where $\Delta{\rm RV}$ is the difference between the measured RV in subsequent epochs and $\Delta t_{\rm RV}$ is the time between the two observations. Because many systems have no significant RV differences, we limited our exploration to systems with $\Delta{\rm RV}>1$\,km s$^{-1}$ \cite{Badenes}. The measured maximum acceleration then gives an estimate for the unseen companion mass of $M({a_{\rm max}})\sim {a}_{\rm max}s^2/G$, where $s$ is the separation between the two bodies over the time  $\Delta t_{\rm RV}$. The separation $s$ is unknown, but in the absence of other information we used $s=\Delta{\rm RV}\,\Delta t_{\rm RV}$ at ${a}={a}_{\rm max}$, which yields an expression similar to the mass function $M({a_{\rm max}})= \Delta{\rm RV}^3\Delta t_{\rm RV}/G$ evaluated for the two epochs for which ${a}={a}_{\rm max}$. This of course is not meant to faithfully return the unseen companion mass, only as a very simple first method to prioritize $10^5$ systems. 

We then acquired ASAS-SN lightcurves \cite{Shappee14,Kochanek17} for the $\sim200$ systems with the highest $M(a_{\rm max})$ as estimated from the APOGEE data. Our hope was to get an estimate for the  orbital period for some systems using the photometric variability expected from ellipsoidal variations, eclipses, or starspots. Many systems showed no variability. These may be interesting for additional followup because in some cases large orbital periods or high inclinations may be implied. Other systems showed periodic photometric variations. 

The system in our sample with the longest well-measured photometric period was 2MASS J05215658+4359220 (``J05215658"). The raw aperture photometry lightcurve from the ASAS-SN Sky Patrol lightcurve server \cite{Kochanek} is shown in Figure \ref{figure:raw}. The phased lightcurve is shown in Figure \ref{figure:lc}.  

We use the Generalized Lomb-Scargle (GLS) \cite{gls,scargle}  to derive the photometric period of the ASAS-SN light curve, finding a best-fit of $P_{\rm phot}\simeq82.2$\,day. To estimate the uncertainty in the period, we calculated the FWHM of the GLS periodogram peak to be $\simeq4.9$\,day. Using the Multi-Harmonic Analysis Of Variance periodogram \cite{Schwarzenberg-Czerny}, we derive a photometric period consistent with that obtained from the GLS periodogram. Although the best-fit photometric period differs from the best fit RV period ($P=83.205\pm0.064$\,day; Table \ref{table:rv}) by $\sim1$\%, the results for the two periods are consistent. 
 
Given the 3 APOGEE measurements listed in Table \ref{table:rv}, we estimated the RV semi-amplitude to be $K\sim(42.6+37.4)/2\sim40$\,km s$^{-1}$. Assuming that the orbital period is equal to the photometric period $P_{\rm orb}=P_{\rm phot}$ (starspots in a tidally locked binary) or that $P_{\rm orb}=2P_{\rm phot}$ (as expected for ellipsoidal variations) we estimated a large value of the mass function $f(M)>0.6$\,M$_\odot$ (eq.\ \ref{mass_function}). Assuming that the observed giant had a mass larger than 1\,M$_\odot$, the implied minimum companion mass is above the Chandrasekhar mass of $\simeq1.4$\,M$_\odot$. Given the absence of any evidence for a stellar companion (Section \ref{section:giant}), we initiated RV followup to measure the orbital period and precise multi-band photometric followup to give both a densely-sampled lightcurve and to constrain potential color variations. 

\begin{figure*}
\centerline{\includegraphics[width=15.5cm]{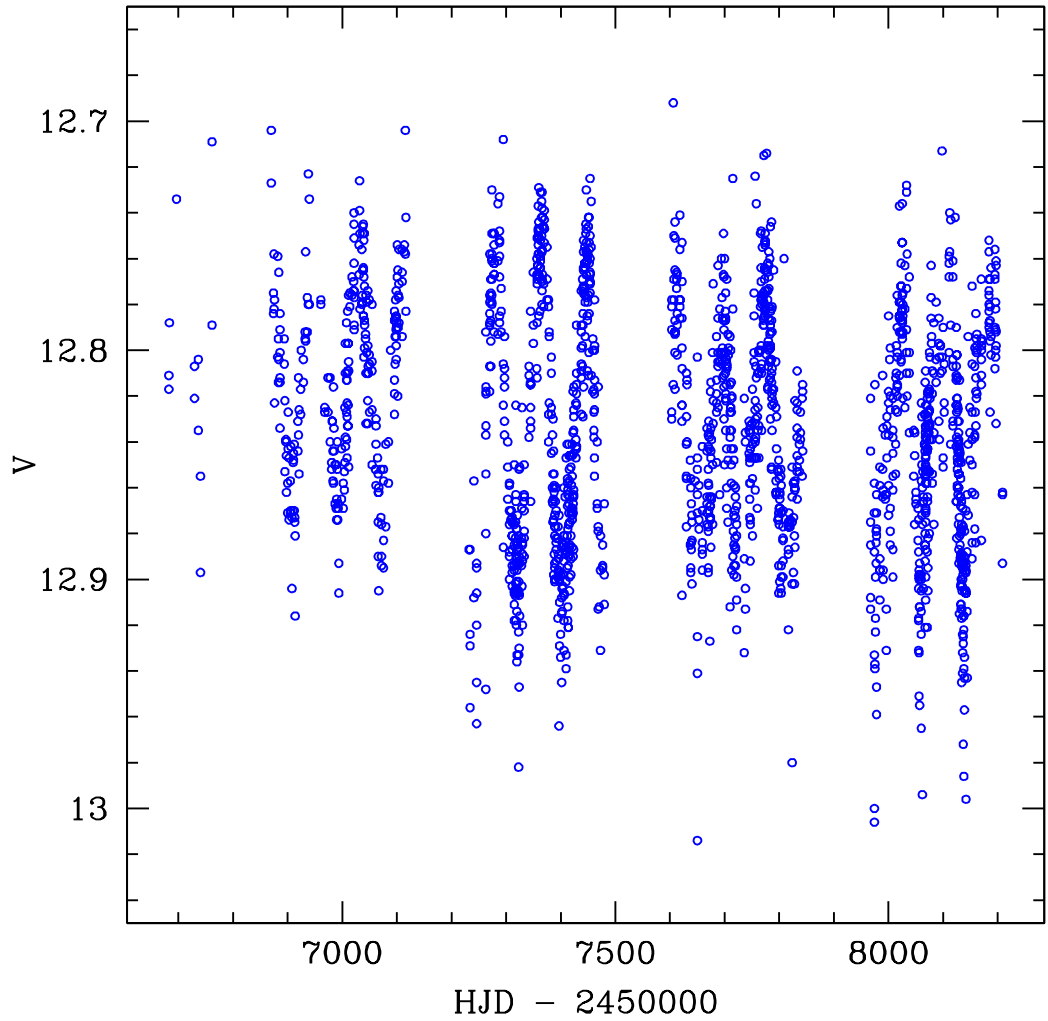}}
\vspace*{-3cm}
\caption{Raw aperture photometry lightcurve of J05215658 in ASAS-SN, taken from the public Sky Patrol lightcurve server \cite{Kochanek17}. The photometric periodicity is evident. Compare with Figure \ref{figure:post} in the main text.}
\label{figure:raw}
\end{figure*}

\section{Multi-Band Optical Photometric Followup}
\label{section:post}

To establish the nature of the photometric variability of J05215658, we obtained multi-band (BV$ri$) images at the Post Observatory Mayhill (NM, USA), which employs a robotic ACP controlled 0.61m CDK telescope with a back illuminated Apogee U47 camera. Between  12 September 2017 and 14 February 2018 (a time span of about 155 days), from 90 to 120 60-sec images were obtained in each band. To extract an instrumental light curve in each filter, we used standard image subtraction procedures \cite{Alard,Hartman}. As our target star is relatively red, the resulting light curves have photon-noise precision significantly better than 1\% in V$ri$-filters and around 1\% in the B-band filter. The resulting lightcurve is shown in the top panel of Figure \ref{figure:post} in the main text. The average calibrated magnitudes derived from these observations are given in Table \ref{table:giant_photometry}.

\section{Gaia Parallax, Binary Motion, and Proper Motion}
\label{section:gaia}

\subsection{Parallax, Offset, and Uncertainty}
The parallax listed in Gaia DR2 for J05215658 (source ID 207628584632757632) is $0.272\pm0.049$, implying a distance of 3.7\,kpc. However, it is known that the parallaxes in Gaia DR2 have a negative zero point, which should be subtracted from the catalog values, thus making the parallaxes larger. This zero point is approximately $-0.030$\,mas for faint quasars, and varies by about $0.043$\,mas (RMS) depending on the position of the object on the sky \cite{Lindegren}. There is also evidence that the zero point varies with magnitude and that it is more negative for bright ($G < 15$) sources. For example, Ref.\ \cite{Riess2018} find an offset of $-0.046\pm 0.013$\,mas for bright Cepheids ($G\sim9$), Ref.\ \cite{Zinn2018,Zinn2017} find an offset of $-0.0528 \pm 0.0024 \pm 0.001$\,mas from a comparison with asteroseismic data for stars in the Kepler field with $G \sim 12$, Ref.\ \cite{Stassun2018} find an offset of $-0.083\pm0.033$\,mas from bright eclipsing binaries ($G\lesssim12$), and Ref.\ \cite{Muraveva2018} find an offset of  $-0.057$\,mas for RR Lyrae stars with $G\sim12$. The variation in the zero point offset for bright sources as a function of position on the sky is not known, but it is reasonable to assume that it is similar to the faint variation: $0.043$\,mas RMS.

Given the Gaia DR2 catalogue value of $0.272 \pm 0.049$\,mas, and assuming a zero point offset of $-0.050$ given the Gaia magnitude $G\simeq12.3$ for J05215658, we take the measured parallax to be 
\begin{equation}
\pi=0.322\,{\rm mas} \pm 0.049\,{\rm mas \,\,(random) \pm 0.043\,mas\,\,(systematic)},
\label{parallax}
\end{equation} 
implying a nominal distance of $D\simeq3.1_{-0.5}^{+0.8}$\,kpc. For comparison, the analysis of  \cite{Bailer-Jones} gives  a distance of $D\simeq3.3^{+0.6}_{-0.5}$\,kpc.
 
\subsection{Binary Motion and Parallax Bias}
We considered the possibility that the parallax given in equation (\ref{parallax}) could be biased by the orbital motion of the system.  The semi-major axis of the orbit of the giant around the center of mass  is $s = a[M_{\rm CO}/(M_{\rm CO}+M_{\rm giant})]$, where $a$ is the semi-major axis of the relative orbit. Using the observed mass function $f(M)$ and the period $P$,
\beq
s=\left(\frac{G\,f(M)P^2}{4\pi^2\sin^3i}\right)^{1/3}\simeq\frac{0.34\,{\rm AU}}{\sin i}\left(\frac{f(M)}{0.77\,{\rm M}_\odot}\right)^{1/3}\left(\frac{P}{83.2\,{\rm days}}\right)^{2/3},
\eeq
so that the ratio of the angular orbital motion to the parallax is 
\beq
\frac{\rm orbital \,\,motion}{\rm parallax}\simeq\,\frac{0.34}{\sin i}\,\left(\frac{f(M)}{0.77\,{\rm M}_\odot}\right)^{1/3}\left(\frac{P}{83.2\,{\rm days}}\right)^{2/3}.
\label{motion}
\eeq
For a nominal parallax of $\pi\simeq0.322$\,mas (eq.~\ref{parallax}), equation (\ref{motion}) then implies that the binary motion will be $0.34\times0.322\,{\rm mas}\simeq(0.11\,{\rm mas})/\sin i$. 

In order to estimate what effect this motion might have on the measured parallax, we made a Monte Carlo simulation of a circular orbit with period 83.2\,days. The orientation of the orbit in space was taken to be random, while the phase of the motion was constrained according to the RV curve (Figure \ref{figure:post}). We then used the timing and geometric scan angle data from the Gaia Observing Schedule Tool (GOST) file\footnote{See https://gaia.esac.esa.int/gost} for J05215658  to calculate the parallax biases resulting from the binary motion by making a least-squares fit of the five astrometric parameters to the calculated orbital displacement of each transit across the focal plane. The 26 GOST entries cataloging the number of transits span more than 500 days and a wide range of scan angles, but half of the observations occurred in a 5 day timespan with two sets of observations in that period highly clustered in time and scan angle. 

Figure \ref{figure:bias} shows the results for $10^4$ model systems, randomly sampling the unknown angles associated with the system's projection on the sky. 
In the left panel the fractional parallax bias of the single-star astrometric solution caused by the orbital motion of the binary is plotted against $\sin i$ for the $10^4$ systems. The bias scales with the size of the orbit and hence with the parallax according to equation (\ref{motion}), and is therefore expressed as a fraction of the true parallax. Our sign convention for the parallax bias is such that a negative bias means that the true parallax is larger than the measured parallax, whereas a positive parallax bias means that the true parallax is smaller than the measured parallax. We find that there could be a positive or negative parallax bias, and that the fractional size of the effect can be very large $\sim\pm1$ for nearly face-on configurations with $\sin i\sim0$. 

In the right panel of Figure \ref{figure:bias} we convert the fractional bias to units of mas by multiplying by an assumed true parallax of $\pi = 0.322$\,mas (eq.~\ref{parallax}) and produce histograms of the parallax bias for $\sin i>0.0$ (black), $>0.6$ (red), and $>0.8$ (blue). The bias has strong peaks at $\pm0.026$\,mas. For $\sin i>0.8$, the bias is essentially never larger than $\pm0.07$\,mas, whereas for nearly face-on configurations a small percentage of systems could in principle have large biases.  However, as we show in Section \ref{section:ruwe}, such large parallax biases are ruled out by the relatively good fit of the astrometric solution, even without a constraint on $\sin i$.

Cases with positive (negative) parallax bias will make the measured parallax larger (smaller) than the true one, and therefore correspond to the case where the actual distance, stellar radius, and luminosity are greater (smaller) than implied by equation (\ref{parallax}). Because the parallax bias in the left panel of Figure \ref{figure:bias} scales with the \textit{true} parallax, we ask the following question: how large (small) can the \textit{true} parallax reasonably be, such that the bias from the orbital motion, combined with the observational uncertainties in equation (\ref{parallax}), could lead to a \textit{measured} value as small (large) as $0.322$\,mas? This question is answered by Table \ref{table:parallax}, which gives the true parallax values at which a given fraction of the simulated cases results in a measured parallax that is $\le$ or $\ge0.322$\,mas. The fractions in the second column correspond, for a normal distribution, to the number of standard deviations in the first column. Including all $\sin i>0$, the 1-$\sigma$ (68\% confidence interval) parallax is 
\beq
\pi\simeq0.322^{+0.086}_{-0.074}, 
\label{best_parallax}
\eeq
which is the value used in the main text. To be precise, this means that if the true parallax is in fact $0.322+0.086=0.408$ or $0.322-0.074=0.248$, then the probability of obtaining an observed value $\le 0.322$ or $\ge 0.322$, respectively, is $(1-0.68)/2 = 0.16$ (1-$\sigma$). For the 2-$\sigma$ parallax (95\% confidence interval), we find that $\pi\simeq0.322^{+0.224}_{-0.146}$ for all $\sin i>0$. It is not possible to set an upper 3-$\sigma$ limit on the true parallax for $\sin i>0$, because the fractional parallax bias is $\lesssim -1$ in more than $0.13$\% of the cases (see Figure \ref{figure:bias}); these systems could therefore in principle have an arbitrarily large true parallax and still be consistent with the measured small value. As we show in Section \ref{section:ruwe}, these very large values of the parallax bias are ruled out by the astrometric goodness of fit. 

Restricting to more nearly edge-on configurations with $\sin i>0.8$, our determination of the confidence intervals for the parallax are $\pi\simeq0.322^{+0.075}_{-0.069}$ (1-$\sigma$), $\pi\simeq0.322^{+0.155}_{-0.134}$ (2-$\sigma$), and $\pi\simeq0.322^{+0.242}_{-0.199}$ (3-$\sigma$). These numbers are of course subject to our assumption of a zero-point offset of $-0.050$\,mas relative to the reported Gaia DR2 parallax of $0.272$\,mas. 

\begin{table}[!t]
\begin{center}
\caption{Confidence intervals for the parallax relative to $\pi=0.322$\,mas, 
including contributions from statistical and systematic uncertainties (eq.\ \ref{parallax}), and
the binary orbital motion (Figs.\ \ref{figure:bias}, \ref{figure:ruwe}).
\label{table:parallax}}
\begin{tabular}{lcccccc}
deviation & fraction  & $\pi$\,(mas)& $\pi$\,(mas) & $\pi$\,(mas) & $\pi$\,(mas) \\
 & & $\sin i>0.0$& $\sin i>0.2$ &$\sin i>0.6$  &  $\sin i>0.8$ \\
\\
\hline
\hline
\\
$-3$\,$\sigma$. & 0.9787 &0.095  &0.117& 0.122 &0.123\\
$-2$\,$\sigma$  & 0.9772 &0.176  &0.181& 0.186 &0.188\\
$-1$\,$\sigma$ & 0.8413 &0.248  &0.249& 0.252 &0.253\\ 
$+1$\,$\sigma$ & 0.1587 &0.408  &0.406& 0.399 &0.397\\
$+2$\,$\sigma$ & 0.0228 &0.546  &0.527& 0.486 &0.477\\
$+3$\,$\sigma$ & 0.0013 & -- &1.470  &0.588& 0.564\\
\\
\hline
\hline
\end{tabular}
\end{center}
\end{table}

\subsection{Constraints from the Astrometric Goodness of Fit}
\label{section:ruwe}

Our simulations show that a large parallax bias is accompanied by a large increase in the RMS residual to the single-star astrometric solution, which could in principle appear in DR2 as astrometric excess noise or an increased astrometric chi-square. J05215658 has zero excess noise, which nominally means that the single-star model fits the data well; however, the excess noise is problematic in DR2 and a better goodness-of-fit indicator is the Renormalized Unit Weight Error (RUWE) described in  GAIA-C3-TN-LU-LL-124-01\footnote{See https://www.cosmos.esa.int/web/gaia/dr2-known-issues}. 
The RUWE is essentially the square root of the reduced chi-square, and should be around $1.0$ for good single-star astrometric solutions. J05215658 has ${\rm RUWE}= 1.055$, which indicates that additional noise from orbital motion is small. 
 
Because the binary motion has to be relatively small in comparison with the random errors per observation in order to get a RUWE close to unity, the observed RUWE can be used to rule out systems with large contributions to the RMS residual from binary motion. From the GOST data described earlier we find that the stated DR2 parallax uncertainty corresponds to an RMS uncertainty per observation of $0.28$\,mas. The astrometric residuals caused by the orbital motion must be considerably smaller than this, or about $0.1$\,mas, to be consistent with the observed RUWE. Because the residuals caused by the orbital motion scale with the size of the orbit, this in turn sets a limit on the size of the orbit and hence on the parallax.

To quantify this argument we made additional simulations where, for each random system, the expected RUWE was computed in addition to the parallax bias. A sample of results are shown in Figure \ref{figure:ruwe}. Each point shows the computed RUWE and measured parallax for a random simulated system with true parallax equal to $0.3$\,mas (upper left panel), $0.5$\,mas (upper right panel), $0.7$\,mas (lower left panel), and $0.9$\,mas (lower right panel). Consistent with our expectations, we see that a large offset between the measured parallax and the true parallax (the parallax bias) comes with an increase in the value of the RUWE. 

In Figure \ref{figure:ruwe}, each panel is constructed from a sample of $10^4$ modeled systems at the given true parallax. Figure \ref{figure:trueplx} summarizes the results from simulations using a much larger sample to improve the statistics. The left panel of Figure \ref{figure:trueplx} shows, as a function of the true parallax, the fraction of systems that have measured parallax $\le 0.322$\,mas for different limits on $\sin i$. The dashed lines correspond to the fractions at 1-, 2-, and 3-$\sigma$ level in Table \ref{table:parallax}, and their intersections with the curves confirm the upper confidence limits in the table. The right panel shows the fraction of systems with measured parallax $\le 0.322$\,mas and $\text{RUWE}\le 1.1$, i.e.\ consistent with the observed goodness of fit with some margin.  Even without a constraint on $\sin i$ the small observed value of the RUWE for J05215658 rules out a true parallax greater than $\simeq 0.62$\,mas at the 3-$\sigma$ confidence level; for $\sin i>0.6$ or $>0.8$, on the other hand, the RUWE does not significantly affect the limits in Table \ref{table:parallax}.

\subsection{Proper Motion}
Gaia measures proper motions of ${\rm pmra}  = -0.055\pm0.10$\,mas yr$^{-1}$ and ${\rm pmdec} = -3.69\pm0.07$\,mas yr$^{-1}$, implying a total proper motion of $\mu\simeq3.69\pm0.07$\,mas yr$^{-1}$ and a total tangential velocity of $\simeq52.5$\,km s$^{-1}$ $\,(D/3\,{\rm kpc})$, significantly larger than the absolute radial velocity of $3.56\pm 0.1$ \,km s$^{-1}$ from the TRES analysis (Section \ref{section:rv}). Taking into account the correlation coefficient ${\rm pmra\_pmdec\_corr}\, = -0.490$, in Galactic coordinates, these proper motions translate to $\mu_l = 3.01\pm0.10$\,mas yr$^{-1}$ and $\mu_b = -2.13\pm0.07$\,mas yr$^{-1}$. The simulations discussed above using the GOST data to assess the parallax bias show that the proper motion components are not likely to be biased by more than a few tenths of a mas yr$^{-1}$ from the binary orbital motion.

We note that the $J$-band reduced proper motion (RPM) for J05215658 is ${\rm RPM}_J=J+\log (\mu)\simeq-2.33$, where $J\simeq9.83$ is the 2MASS magnitude (Table \ref{table:giant_photometry}) and $\mu$ is the total proper motion in arcseconds per year. With a color of $J-H\simeq0.76$, the system falls among the giants in a reduced proper motion diagram \cite{Stassun2018RPM}.  We also used the Gaia Universe Model Snapshot (GUMS) simulation from Ref.\ \cite{gums}  to compare the observed proper motion of J05215658 with the expected motions of stars at various distances from the Sun. Extracting sources in GUMS within 5 degrees of J05215658, and with similar apparent $G$ magnitude ($\pm 1$\,mag) and color index $G_\text{BP}-G_\text{RP}$ ($\pm 0.25$\,mag), we find that the observed proper motion is typical for a giant at a distance of $1.5-5$\,kpc from the Sun, but improbably small for a more nearby dwarf.

\begin{figure*}
\centerline{\includegraphics[width=8.cm]{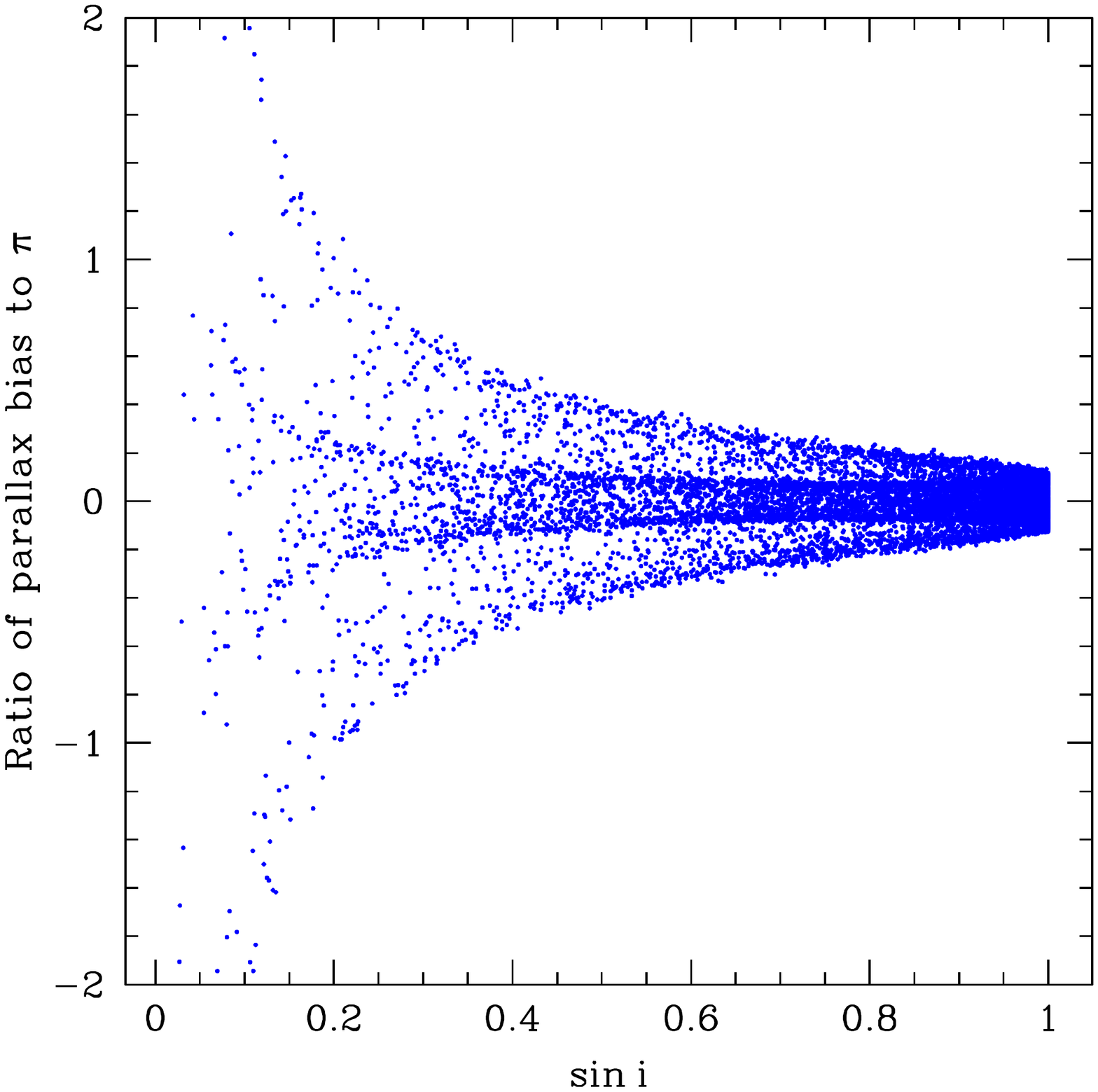}\includegraphics[width=8.cm]{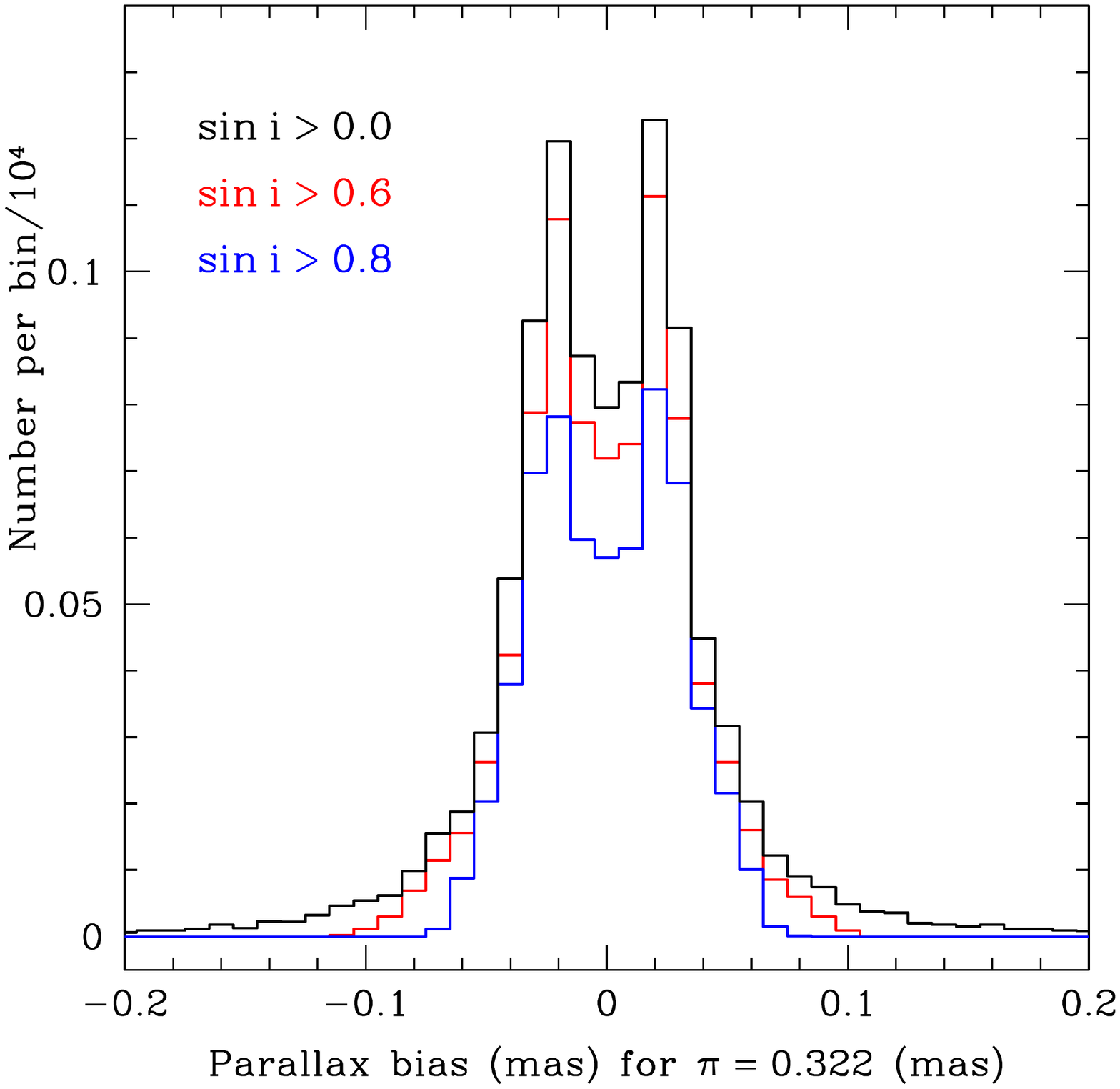}}
\vspace*{-1.75cm}
\caption{Results for the parallax bias of the single-star astrometric solution as a result of the binary motion for J05215658. Left panel: fractional parallax bias as a function of $\sin i$ for $10^4$ simulated systems (see Section \ref{section:gaia}; Table \ref{table:parallax}). Right panel: histograms of the number of simulated systems per bin versus the parallax bias in mas, assuming a true parallax of $\pi=0.322$\,mas (eq.~\ref{parallax}), for $\sin i>0.0$ (black) $>0.6$ (red), and $>0.8$ (blue). The parallax bias is typically $\simeq\pm0.026$\,mas, and less than $\pm 0.07$\,mas for $\sin i > 0.8$.}
\label{figure:bias}
\end{figure*}

\begin{figure*}
\centerline{\includegraphics[width=7.cm]{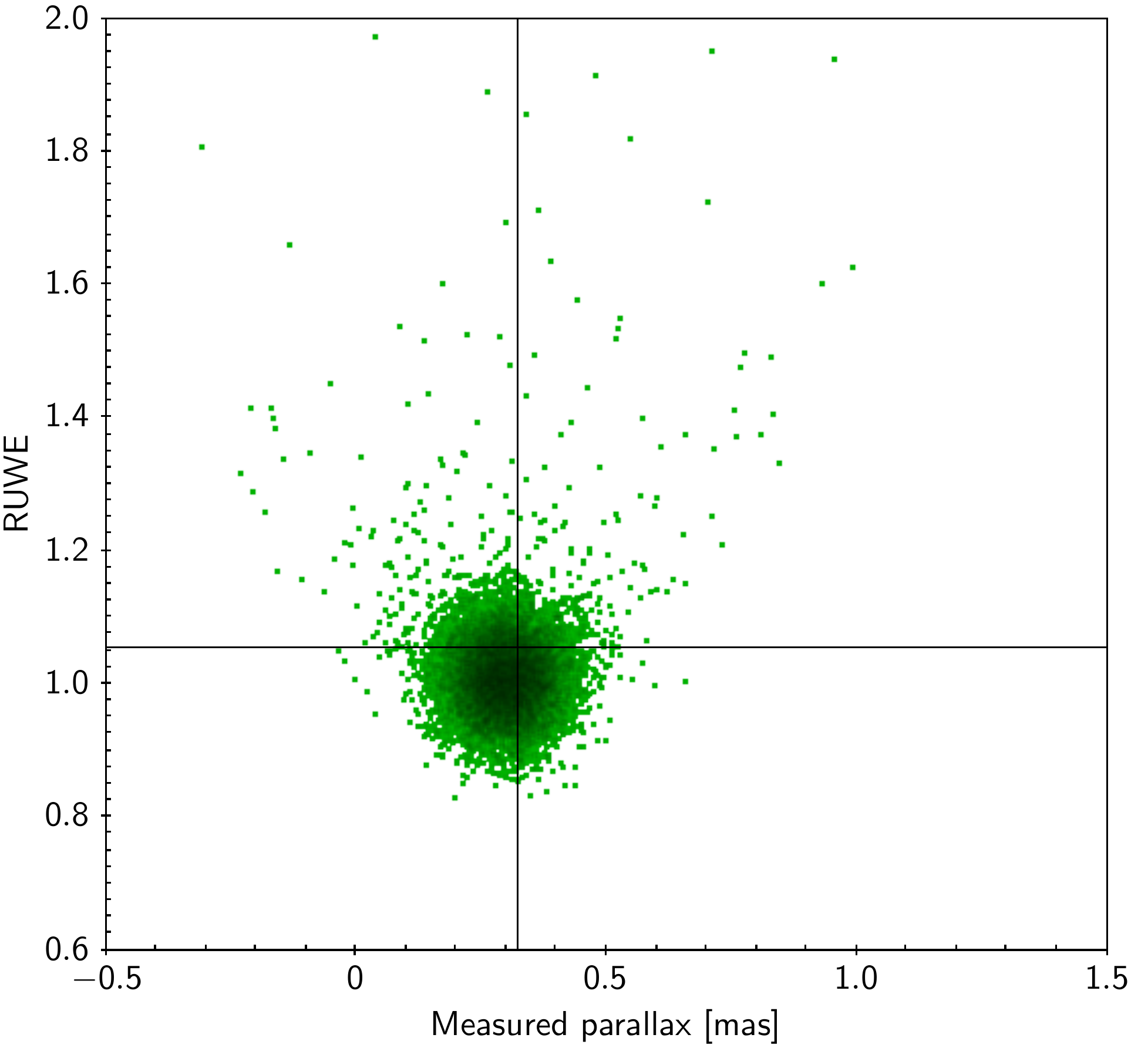}\includegraphics[width=7.cm]{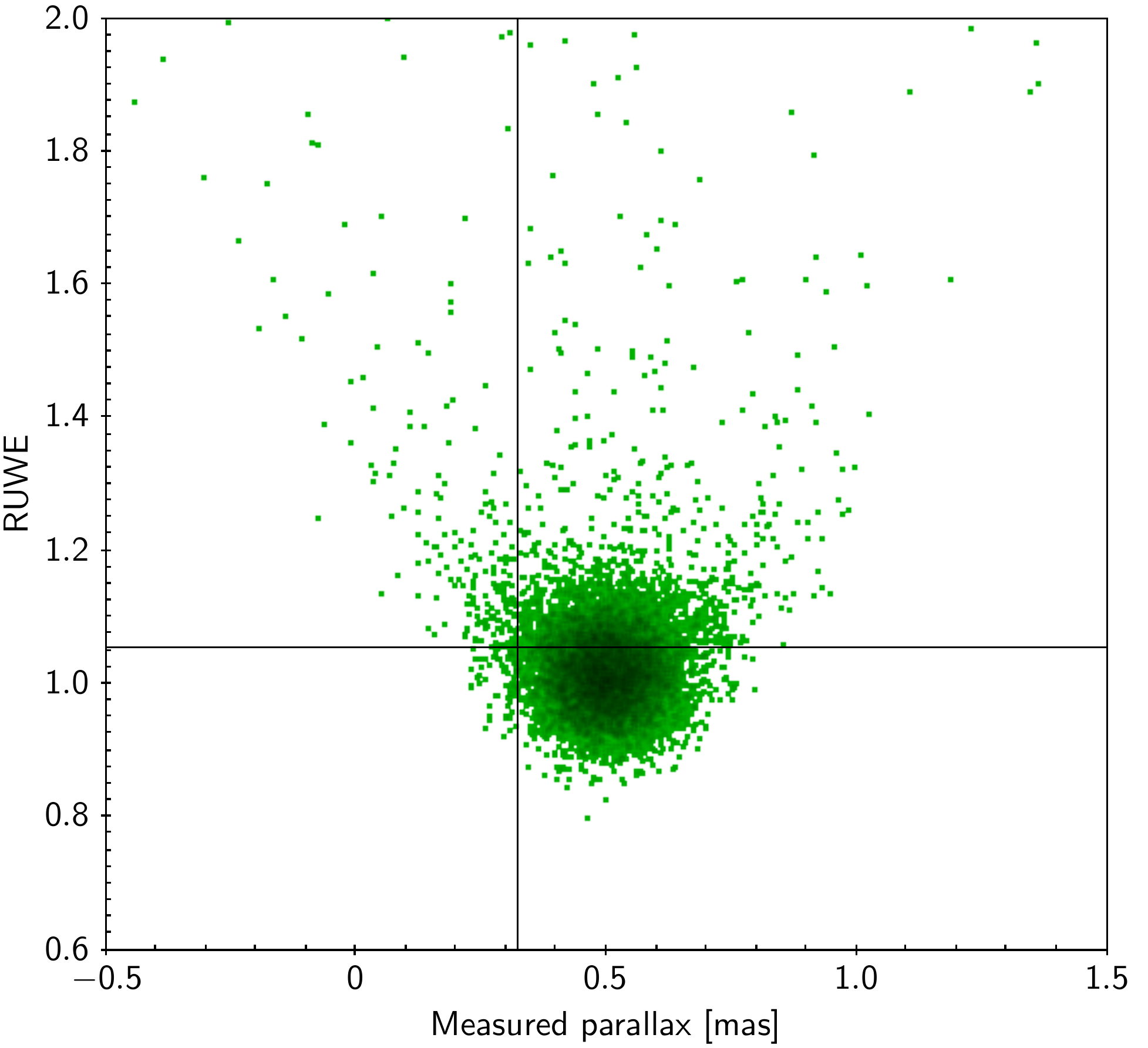}}
\centerline{\includegraphics[width=7.cm]{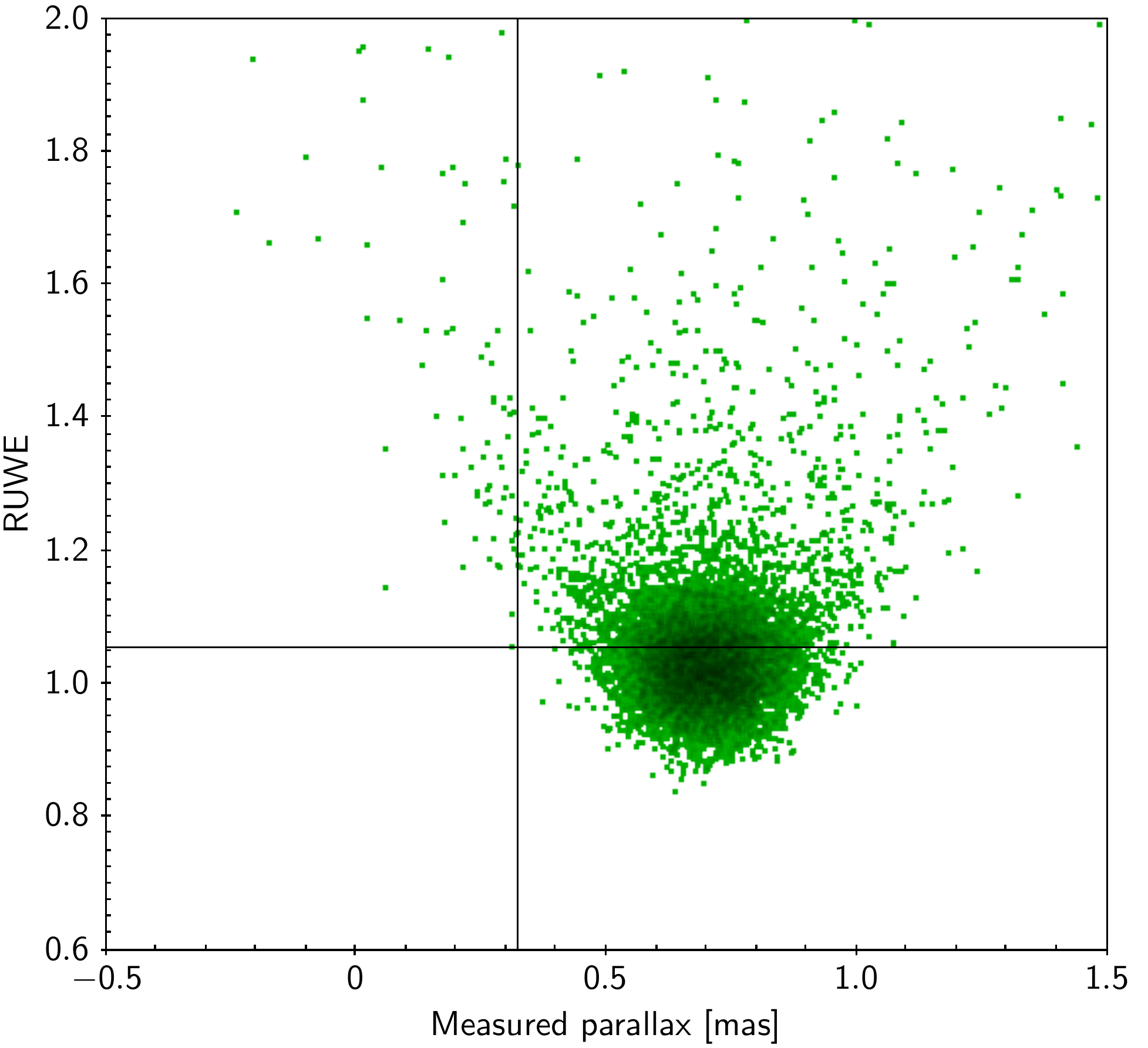}\includegraphics[width=7.cm]{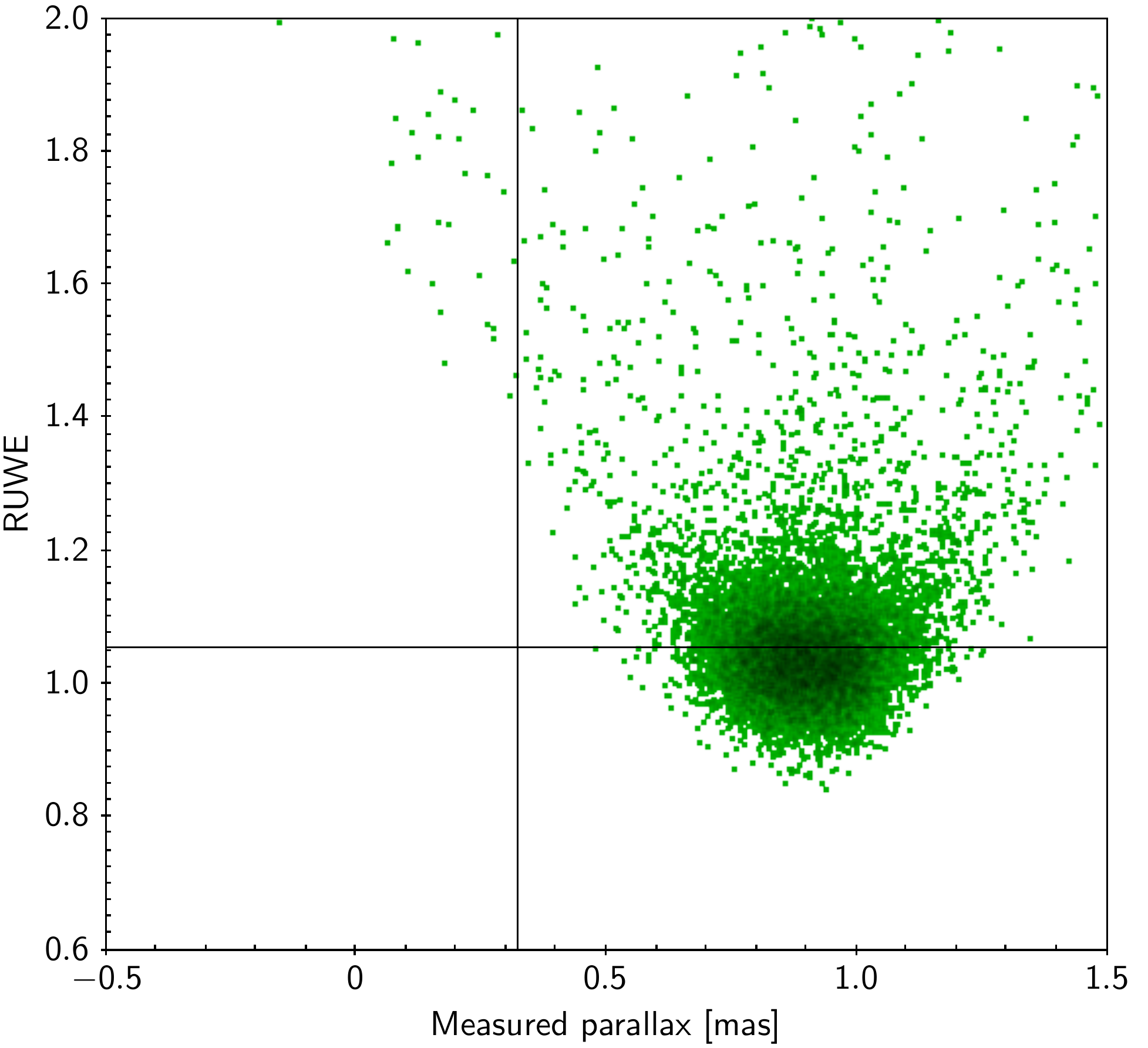}}
\caption{Computed RUWE for model systems of J05215658, including binary motion, as a function of the {\it measured} parallax for an assumed {\it true} parallax of 0.3\,mas (upper left), 0.5\,mas (upper right), 0.7\,mas (lower left), and 0.9 (lower right). Solid lines mark the measured RUWE\,$ =1.055$ and measured parallax of $0.322$\,mas (eq.~\ref{best_parallax}). Because a large absolute value of the parallax bias is accompanied by a large increase in the RMS residual to the single-star astrometric solution (Fig.\ \ref{figure:bias}), the low value of the measured RUWE rules out the possibility that the true parallax could be $\gtrsim0.62$\,mas (see Figure \ref{figure:trueplx}).}
\label{figure:ruwe}
\end{figure*}

\begin{figure*}
\centerline{\includegraphics[width=8.cm]{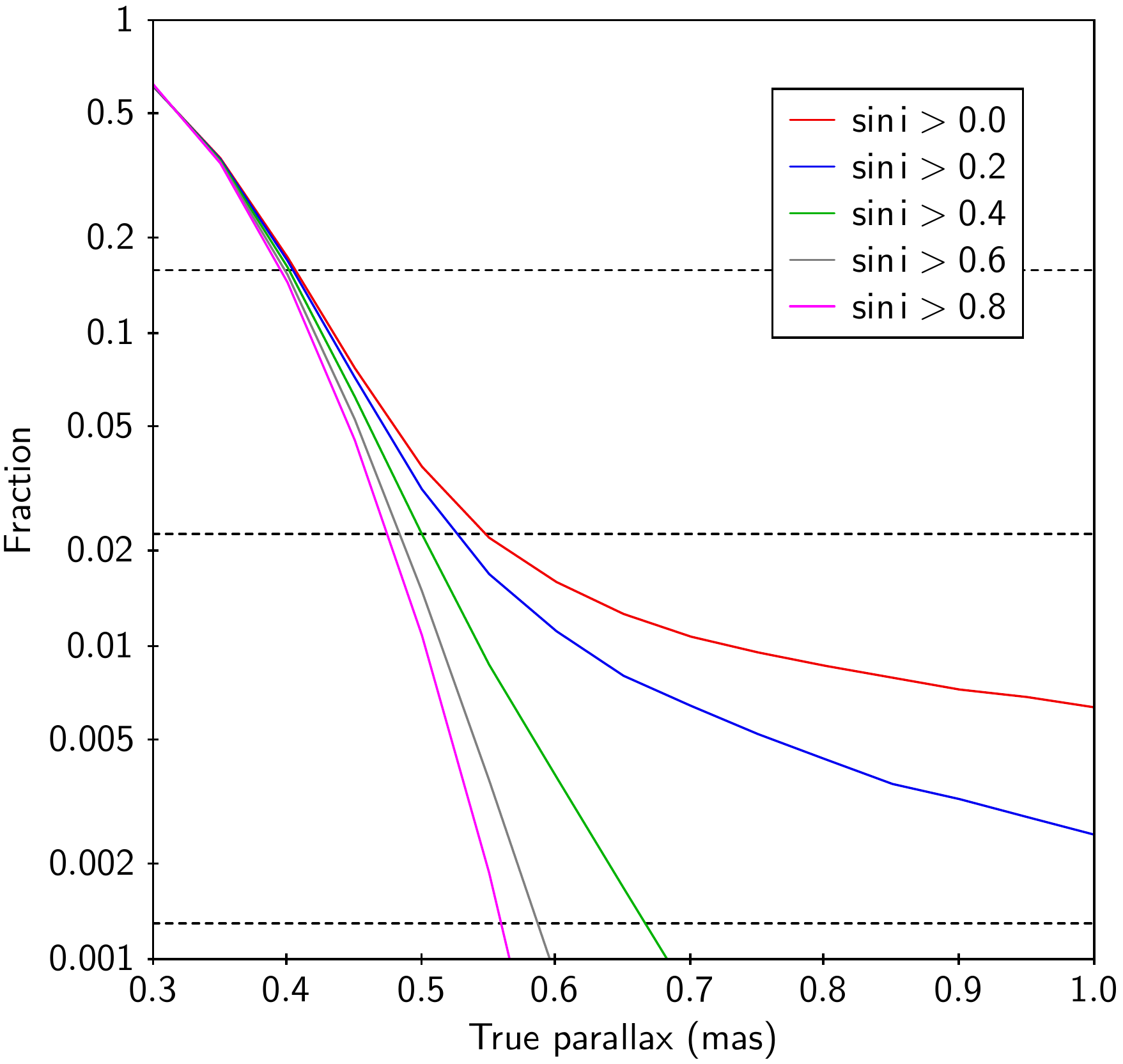}\includegraphics[width=8.cm]{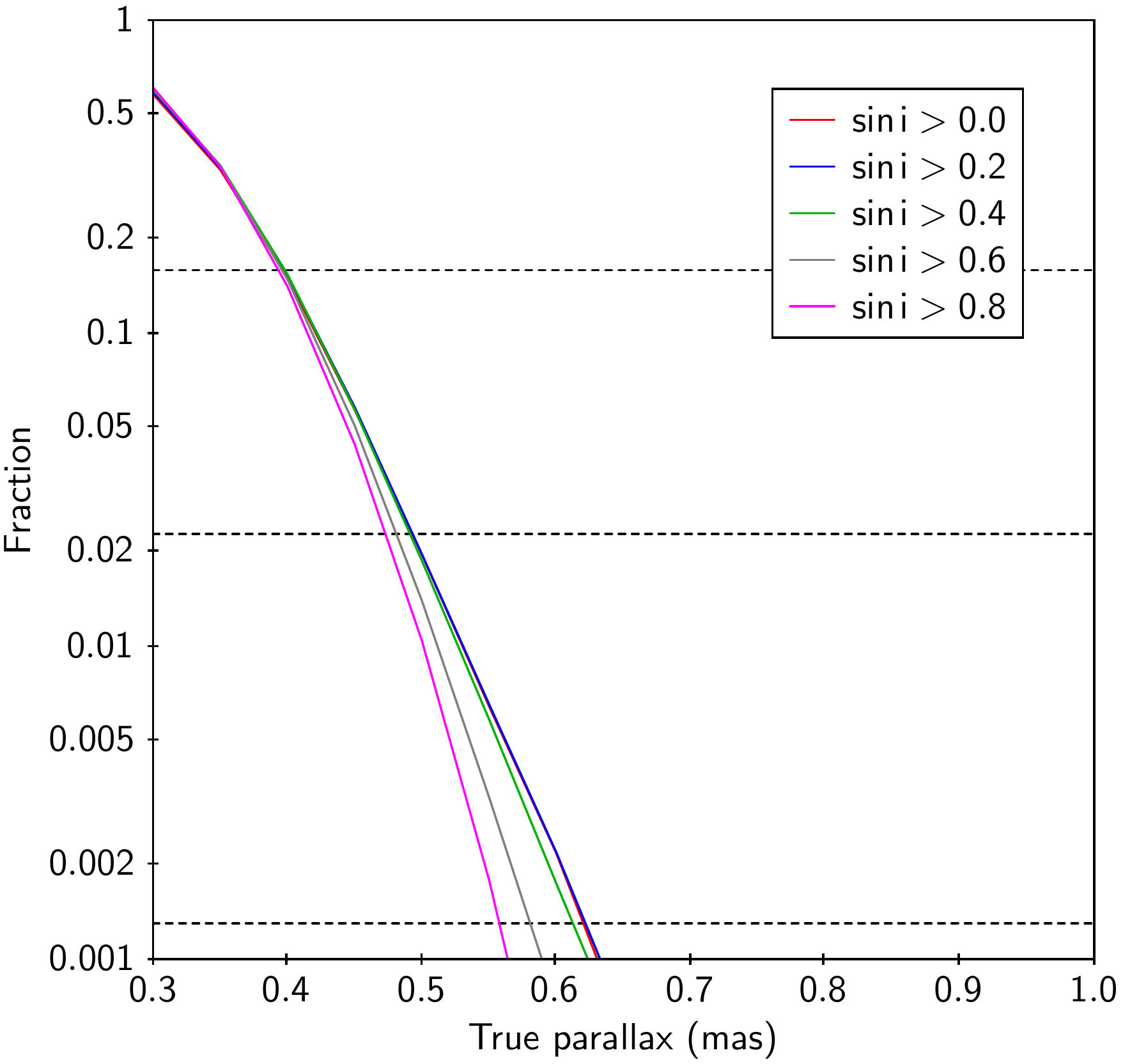}}
\caption{Fraction of model systems for which the computed parallax and astrometric goodness of fit are consistent with observed values, as a function of the assumed true parallax of the system. Left panel: fraction of systems with measured parallax $\le 0.322$\,mas. Right panel: fraction of systems with measured parallax $\le 0.322$\,mas and $\text{RUWE}\le 1.1$. The curves are for $\sin i > 0$ (red), $>0.2$ (blue), $>0.4$ (green), $>0.6$ (gray), and $>0.8$ (magenta). The horizontal dashed lines show the fractions corresponding to the upper 1-, 2, and 3-$\sigma$ confidence limits in Table \ref{table:parallax}. The nearly face-on cases (small $\sin i$) in the left panel consistent with a large true parallax are in the right diagram ruled out by the observed RUWE.}
\label{figure:trueplx} 
\end{figure*}

\section{Radial Velocity Followup}
\label{section:rv}

We initiated spectroscopy with the Tillinghast Reflector Echelle Spectrograph (TRES; \cite{fur}) on the 1.5m Tillinghast Reflector at the Fred Lawrence Whipple Observatory (FLWO) located on Mt. Hopkins in Arizona. TRES has a resolution of $R \sim 44,000$ and spectra were collected using the medium 2.3" fiber.
 
A total of 11 spectra were obtained between 10 September 2017 and 25 January 2018. The spectra were reduced and extracted as described in \cite{buchhave10}. The exposure times ranged from $30-42$ minutes depending on observing conditions and yielded an average signal-to-noise per resolution element (SNRe) of $\sim25$ at the peak of the continuum centered at 519\,nm surrounding the Mg b triplet. We derived relative radial velocities using the observation with the highest SNRe as a template and cross-correlated the remaining spectra order-by-order against the observed template. The multi-order velocity analysis avoids including orders that have significant contamination by telluric lines.  The errors for the multi-order velocities are based on the velocity scatter, order by order. The observed template is, by definition, assigned a velocity of $0$\,km s$^{-1}$. 
 
The absolute velocity of the system is $3.56\pm 0.1$ \,km s$^{-1}$. This is derived from the radial velocity for the template observation when correlated against our library of calculated spectra using the Mg b order, combined with a $-0.61$\,km s$^{-1}$ correction, which is mostly due to the fact that the calculated template spectrum does not include gravitational redshift. The uncertainty is based on residual systematics of many years of observations of the International Astronomical Union (IAU) Radial Velocity Standard Stars. The derived relative radial velocity results from TRES are given in Table \ref{table:rv} and the bottom panel of Figure \ref{figure:post}. Note that the absolute APOGEE radial velocities listed in Table \ref{table:rv} were not included in the fit reported in Table \ref{table:orbital}. We experimented with including the APOGEE data in the RV fit using the code radvel \cite{radvel}, and found results consistent with those reported in Table \ref{table:orbital}, but since the APOGEE data likely have significantly larger systematic uncertainties than the $\sim10$\,m s$^{-1}$ reported in Table \ref{table:rv} \cite{Badenes}, and since there are likely additional systematic offsets between the two instruments, we prefer to report  the derived relative radial velocity results from TRES alone.

\begin{table}[!t]
\begin{center}
\caption{RV Measurements from APOGEE and TRES.
\label{table:rv}}
\begin{tabular}{lccc}
\hline \hline
\\
APOGEE:\\
\\
\multicolumn{1}{c}{JD} &
\multicolumn{1}{c}{Absolute RV} &
\multicolumn{1}{c}{Uncertainty}\\
\multicolumn{1}{c}{$-2450000$} &
\multicolumn{1}{c}{(km/s)} &
\multicolumn{1}{c} {(km/s)}\\
\\
$6204.9537$  & $-37.417$  & $0.011$ \\
$6229.9213$  & $34.846$   & $0.010$\\
$6233.8732$  & $42.567 $  & $0.010$ \\
\\
\hline
\\
TRES:\\
\\
\multicolumn{1}{c}{BJD} &
\multicolumn{1}{c}{Relative RV} &
\multicolumn{1}{c}{Uncertainty}\\
\multicolumn{1}{c}{$-2450000$} &
\multicolumn{1}{c}{(km/s)} &
\multicolumn{1}{c} {(km/s)}\\
\\

$8006.9760 $  &  $    0.000  $  &  $   0.075    $ \\
$8023.9823 $  &  $ -43.313  $  &  $   0.075 $ \\
$8039.9004 $  &  $ -27.963  $  &  $   0.045 $ \\
$8051.9851 $  &  $  10.928  $  &  $   0.118$ \\
$8070.9964 $  &  $  43.782  $  &  $   0.075   $ \\
$8099.8073 $  &  $ -30.033  $  &  $   0.054 $ \\  
$8106.9178 $  &  $ -42.872  $  &  $   0.135 $ \\ 
$8112.8188 $  &  $ -44.863  $  &  $   0.088  $ \\ 
$8123.7971 $  &  $ -25.810  $  &  $   0.115  $ \\
$8136.6004 $  &  $  15.691  $  &  $   0.146  $ \\
$8143.7844  $  &  $ 34.281  $  &  $   0.087 $ \\  
\\
\hline
\hline
\end{tabular}
\end{center}
\end{table}

\begin{table}[!t]
\begin{center}
\caption{Orbital Parameters derived from TRES RV measurements listed in Table \ref{table:rv}.
\label{table:orbital}}
\begin{tabular}{lccc}
\hline
\hline
\\
$P$ & $83.205 \pm 0.064$ &days\\
$T$ & $58115.93\pm7.4$  & BJD$-2400000$\\
$e$ & $0.00476\pm0.00255$ & \\
$\omega$ & $197.13\pm32.07$ &degrees \\
$K$ & $44.615\pm0.123$ &km s$^{-1}$\\
$\gamma$ & $-0.389 \pm 0.101$ &km s$^{-1}$ \\
$f(M)$ & $0.766\pm0.00637$ &M$_\odot$ \\
\hline
\hline
\end{tabular}
\end{center}
\end{table}

\section{Properties of the Giant}
\label{section:giant}

Archival and new photometry of the system is summarized in Table \ref{table:giant_photometry}. In addition to the data we collected as part of our multi-band photometric followup, we obtained Swift imaging, which yielded a detection in the U and UVM2 bands and an upper limit in the X-ray. The UVM2 detection is important for constraining a stellar companion, as discussed in Section \ref{section:limits}. The X-ray upper limit is discussed in Section \ref{section:x}.

\subsection{Analysis of TRES Spectra}
\label{section:tres_star}

We used the Stellar Parameter Classification (SPC) tool to derive stellar parameters from the TRES spectra \cite{buchhave12} discussed in Section \ref{section:rv}. SPC cross correlates the observed spectra against a library of synthetic spectra based on Kurucz model atmospheres. We ran SPC on each TRES spectrum individually and then report the results as a weighted average. The weighted average results from this analysis are $T_{\rm eff}=4574\pm 65$\,K, $\log g =2.35\pm 0.14$, and $[{\rm m/H}]=-0.39\pm 0.08$. The total line broadening parameter is found to be $V_{\rm broad}=16.8\pm 0.6$\,km s$^{-1}$, which accounts for the instrumental broadening of TRES ($6.8$\,km s$^{-1}$), but which does not distinguish between the contributions from rotational broadening and macroturbulence. 

Ref.\ \cite{Carney2008b} (see also Ref.\ \cite{Massarotti}) suggests an empirical fit to results from high-resolution spectroscopic data to connect the total line broadening  parameter and its contributions from rotation and macroturbulence for giants of the form
\beq
V_{\rm broad}=\left[(v \sin i)^2+0.95\xi_{\rm RT}^2\right]^{1/2},
\label{xirt}
\eeq
where $\xi_{\rm RT}$ is the unknown radial-tangential macroturbulent dispersion. Using this relation, in Table \ref{table:xirt}  we report the implied value of $v \sin i$ for values of $\xi_{\rm RT}$ ranging from $0-10$\,km s$^{-1}$.\footnote{Ref.\ \cite{Carney2008b} also provides an empirical relation between the total broadening parameter and a higher resolution determination of $v \sin i$ (their equation 2) for their sample of both red giant branch and red horizontal branch stars. Directly applying their relation with the TRES determination of $V_{\rm broad}$, we obtain $v\sin i\simeq13.4$\,km s$^{-1}$. Given the quoted scatter of 1.5\,km s$^{-1}$ in this relation, this determination of $v\sin i$ is in agreement with the APOGEE $v \sin i$ (see Section \ref{section:apogee}) and with the range of $\xi_{\rm RT}$ considered in Table \ref{xirt}.}  No red giant in the sample of Ref.\ \cite{Carney2008b} has $\xi_{\rm RT}$ greater than $\simeq10$\,km s$^{-1}$, and those that have $\xi_{\rm RT}\simeq10$\,km s$^{-1}$ are significantly more luminous than J05215658 for the nominal Gaia parallax (see Section \ref{section:gaia}).  The set of giants with properties closest to J05215658 in Ref.\ \cite{Carney2008b} have $\xi_{\rm RT}\sim 4-7$\,km s$^{-1}$. Indeed, the empirical fits to $\xi_{\rm RT}$ summarized in equations 4, 5, and 6, of Ref.\ \cite{Carney2008b} give estimates for $\xi_{\rm RT}$ of $5.1$, $5.5$, and $5.6$\,km s$^{-1}$ for J05215658.\footnote{Note that Ref.\ \cite{Carney2008b}'s equation 6 should be modified with the substitution $T_{\rm eff}\rightarrow\Theta_{\rm eff}$ for consistency with their Figure 10.} These numbers should be compared with the range used for $\xi_{\rm RT}$ in Table \ref{table:xirt}. Assuming a tidally circularized and synchronized system, as in the main text, we use the TRES measurements to derive the minimum giant radius and luminosity, distance and parallax, and mass from $\log g\simeq2.35\pm0.14$ ($M_{\rm giant}^{\log g}=g\,R^2/G$).  Values of all parameters are reported in the Table. Similar estimates are made for the APOGEE $v \sin i$ in Section \ref{section:apogee} and Table \ref{table:xirt_apogee}.

For $\xi_{\rm RT}=0-8$\,km s$^{-1}$, Table \ref{table:xirt} shows that the implied parallax from $v\sin i$ is $\sim0.343 - 0.387$, in good agreement with $\pi=0.322^{+0.086}_{-0.074}$ from Gaia (eq.~\ref{best_parallax}; Table \ref{table:parallax}). For the largest values of $\xi_{\rm RT}$ considered in Table \ref{table:xirt}, the minimum mass inferred from $\log g$ is $\gtrsim3.0$\,M$_\odot$, implying a companion of $M_{\rm CO}\gtrsim3.0$\,M$_\odot$ (Figure \ref{figure:m}). The value of $\xi_{\rm RT}$ required for $M_{\rm giant}^{\log g}$ to equal $1$\,M$_\odot$, as implied by the near-Solar $[{\rm C/N}]$ abundance (see main text and Section \ref{section:cn}), is  $\xi_{\rm RT}\simeq15.8$\,km s$^{-1}$ (for $\log g=2.35$). 

\begin{table}[!t]
\begin{center}
\caption{Properties of the giant using the $T_{\rm eff}$, $\log g$, and $V_{\rm broad}$ derived from the  TRES spectroscopy and assuming different values of $\xi_{\rm RT}$, equation (\ref{xirt}), and that the binary is tidally synchronized. The quantity $M_{\rm giant}^{\log g}=g\,R^2/G$.
\label{table:xirt}}
\begin{tabular}{lccccccc}
$\xi_{\rm RT}$ & $v\sin i$ & $R\,\sin i$ &$L\,\sin^2i$ & $D\,\sin i$ & parallax\,$/\sin i$ & $M^{\log g}_{\rm giant}\,\sin^2 i$ \\
(km s$^{-1}$) & (km s$^{-1}$) & ($R_\odot$) &  ($L_\odot$) & (kpc) & (mas) &  ($M_\odot$) \\
\\
\hline
\hline
\\
0.00    &16.8    &27.6    &301    &2.92    &0.343    &6.23    $^{+2.37    }_{-1.72    }$ \\
2.00    &16.7    &27.5    &297    &2.90    &0.345    &6.15    $^{+2.34    }_{-1.69    }$ \\
4.00    &16.3    &26.9    &285    &2.84    &0.352    &5.90    $^{+2.24    }_{-1.62    }$ \\
6.00    &15.7    &25.9    &264    &2.74    &0.365    &5.48    $^{+2.08    }_{-1.51    }$ \\
8.00    &14.9    &24.5    &236    &2.59    &0.387    &4.89    $^{+1.86    }_{-1.35    }$ \\
10.0    &13.7    &22.5    &200    &2.38    &0.421    &4.13    $^{+1.57    }_{-1.14    }$ \\
\\
\hline
\hline
\end{tabular}
\end{center}
\end{table}

\subsection{Analysis of APOGEE Spectra}
\label{section:apogee}

Analysis of J05215658 using the APOGEE Stellar Parameter and Chemical Abundances Pipeline (ASPCAP) \cite{Perez} yields stellar parameters of $T_{\rm eff}\simeq4480.0\pm62.3$\,K, $\log g\simeq2.59\pm0.06$,   $[{\rm M/H}]\simeq-0.298\pm0.03$,  $[{\rm \alpha/M}]\simeq-0.04\pm0.015$, and  $[{\rm C/N}]\simeq0.0$.\footnote{The ASPCAP analysis data page is available here \url{https://dr14.sdss.org/infrared/spectrum/view/stars=aspcap?id=23413&index=0}.}  The Cannon analysis \cite{Cannon} of the spectra gives similar results: $T_{\rm eff}\simeq4406.4\pm57$\,K, $\log g\simeq2.653\pm0.136$,   $[{\rm M/H}]\simeq-0.309\pm0.059$,  $[{\rm \alpha/H}]\simeq-0.052\pm0.043$. 

We employed the analysis technique used by Ref \cite{Tayar} to determine the projected rotational velocity of the giant in J05215658 from the APOGEE spectra. Figure \ref{figure:vsini} shows a piece of the APOGEE spectrum (black), as well as model spectra including macroturbulence, broadened with $v\sin i=0.0$ (red), 5.0 (green), and 14.1 (blue) km s$^{-1}$ as well as the residuals. Using this method over the full APOGEE spectral range, we find $v \sin i =14.1\pm0.6$\,km s$^{-1}$. This estimate for $v \sin i$ includes macroturbulence through the fitting function proposed by \cite{Holtzman2018}, which gives a macroturbulent broadening parameter of $\simeq3.6$\,km s$^{-1}$. For stars like J05215658, the distribution of macroturbulent broadening parameters is tightly clustered to a sequence near $4$\,km $s^{-1}$, but with some outliers (Fig.\ 9 of Ref.\ \cite{Holtzman2018}). 

As in the discussion of the TRES spectroscopy (see Section \ref{section:rv}), we considered the possibility that the macroturbulence was underestimated for J05215658.  Given the uncertainties in the macroturbulent broadening parameter, we view the TRES and APOGEE $v \sin i$ determinations as reasonably consistent (i.e., TRES: $[16.8^2-0.95\xi_{\rm RT}^2]^{1/2}$ versus APOGEE: $14.1$\,km s$^{-1}$). Still, for completeness, and as a guide to how an underestimate in $\xi_{\rm RT}$ might change the results, we used equation (\ref{xirt}) to estimate the value of the total broadening parameter if the assumed $3.6$\,km s$^{-1}$ had not been subtracted, yielding $V_{\rm broad}\simeq14.5$\,km s$^{-1}$. Then, as in Section \ref{section:tres_star}, we used equation (\ref{xirt}) to estimate the change in our derived minimum stellar radius and luminosity using values of $\xi_{\rm RT}$ ranging from $0-10$\,km s$^{-1}$ \cite{Carney2008b}. Table \ref{table:xirt_apogee} shows the results for the parameters derived. The mass determined from $\log g$ is computed using the TRES $\log g\simeq2.35\pm0.14$ since the APOGEE $\log g$ is likely significantly biased (see below; Fig.\ref{figure:rot}). We find that for $\xi_{\rm RT}$ as large as $10$\,km s$^{-1}$, the implied giant mass is larger than $1.8$\,M$_\odot$, implying a minimum mass for the companion of $M_{\rm CO}\gtrsim2.4$\,M$_\odot$ (Fig.\ \ref{figure:m}). 

The values of $v\sin i$  and $\xi_{\rm RT}$ also enter our comparison to single-star evolutionary models in Section \ref{section:tracks} because once a best-fit stellar radius is determined by comparing to the models, a value of $\sin i$ can be inferred to derive the range of acceptable companion masses.

\begin{table}[!t]
\begin{center}
\caption{Properties of the giant using $V_{\rm broad}$ derived from the  APOGEE spectroscopy, $T_{\rm eff}$ from the SED (Section \ref{section:sed}), and assuming different values of $\xi_{\rm RT}$, equation (\ref{xirt}), and that the binary is tidally synchronized. The quantity $M_{\rm giant}^{\log g}=g\,R^2/G$ uses the TRES value of $\log g$ since the APOGEE $\log g$ measurement is likely biased (see Section \ref{section:apogee}).
\label{table:xirt_apogee}}
\begin{tabular}{lccccccc}
$\xi_{\rm RT}$ & $v\sin i$ & $R\,\sin i$ &$L\,\sin^2i$ & $D\,\sin i$ & parallax\,$/\sin i$ & $M^{\log g}_{\rm giant}\,\sin^2 i$ \\
(km s$^{-1}$) & (km s$^{-1}$) & ($R_\odot$) &  ($L_\odot$) & (kpc) & (mas) &  ($M_\odot$) \\
\\
\hline
\hline
\\
0.00    &14.5    &23.9    &216    &2.47    &.405    &4.64    $^{+1.77    }_{-1.28    }$ \\
2.00    &14.4    &23.6    &212    &2.45    &.408    &4.56    $^{+1.73    }_{-1.26    }$ \\
4.00    &14.0    &23.0    &200    &2.38    &.420    &4.31    $^{+1.64    }_{-1.19    }$ \\
6.00    &13.3    &21.8    &181    &2.26    &.442    &3.89    $^{+1.48    }_{-1.07    }$ \\
8.00    &12.2    &20.1    &153    &2.08    &.480    &3.30    $^{+1.26    }_{-.909    }$ \\
10.0    &10.7    &17.7    &118    &1.83    &.547    &2.54    $^{+.968    }_{-.701    }$ \\
\\
\hline
\hline
\end{tabular}
\end{center}
\end{table}

\begin{figure*}
\centerline{\includegraphics[width=18.cm]{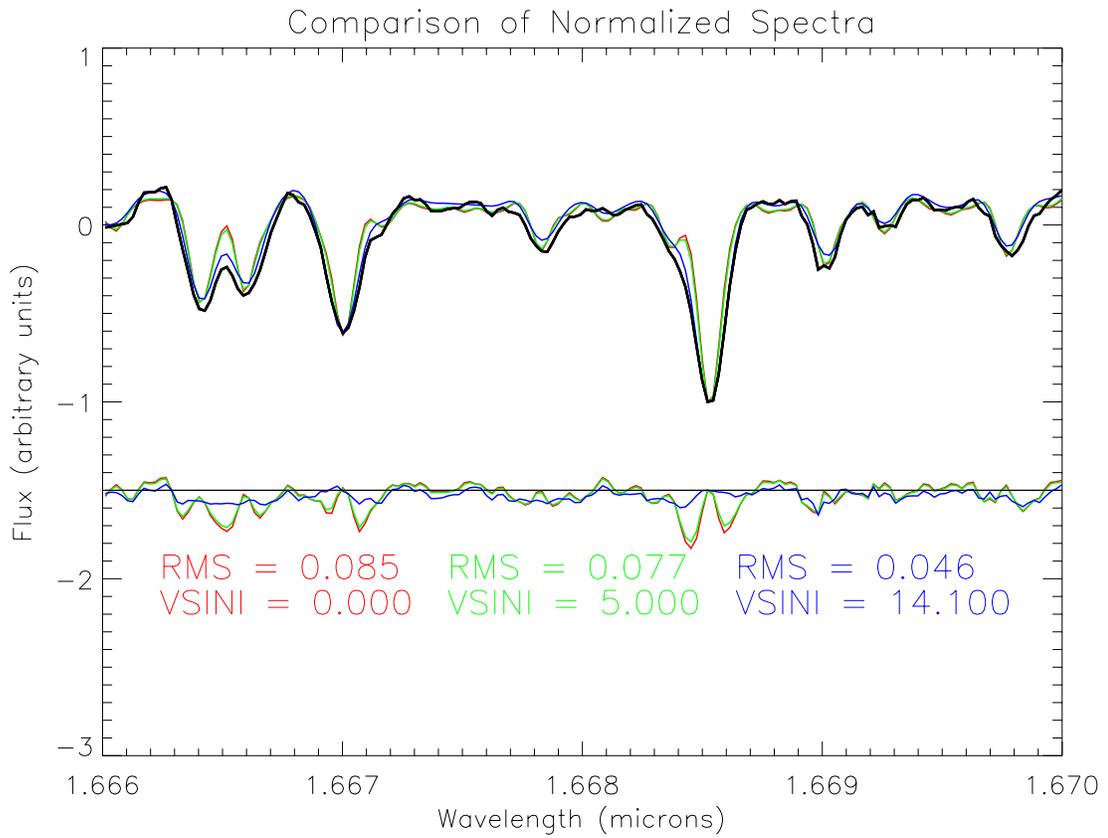}}
\vspace*{-1cm}
\caption{Example of the change in the fit to the observed spectral lines in APOGEE (black) as the projected rotation velocity is increased from $v\sin i=0$ (red), 5 (green), and 14.1\,km s$^{-1}$ (blue). The bottom set of lines shows the residuals with respect to the data. We find a best fit of $v\sin i=14.1\pm0.6$\,km s$^{-1}$. Corrections for macroturbulent broadening are discussed in Section \ref{section:apogee} and Table \ref{table:xirt_apogee}.}
\label{figure:vsini}
\end{figure*}

{\bf APOGEE $\log g$:} The large value of $\log g\simeq2.59\pm0.06$ found by APOGEE ASPCAP DR14 is significantly different than the value obtained from the optical TRES spectra, $\log g\simeq 2.35\pm 0.14$. In order to address this question we compared spectroscopic $\log g$ determinations from APOGEE with those determined from astroseismology from the Apache Point Observatory-{\it Kepler} Asteroseismology Science Consortium (APOKASC)  sample \cite{Pinsonneault18}. We were particularly interested in whether or not rapidly rotating giants might have systematically discrepant spectroscopic $\log g$ as a function of $v\sin i$ \cite{Holtzman2018}. Figure \ref{figure:rot} shows the difference between the spectroscopic $\log g$ and the astroseismic $\log g$ from the APOKASC sample for systems with well-measured $v \sin i$ \cite{Tayar} as a function of $v\sin i$ (upper left), spectroscopic $\log g$ (APOGEE) (upper right), spectroscopic $T_{\rm eff}$ (lower left), and $[{\rm Fe/H}]$ (lower right). The data points are represented with circles whose size is proportional to $v\sin i$ for clarity. We find that the large majority of the data are above 0, indicating that the spectroscopic $\log g$ determination is systematically larger than the astroseismic determination. There is additionally a trend with $v\sin i$ such that the difference between the spectroscopic and astroseismic $\log g$ determinations increases with $v\sin i$. For $v \sin i \simeq8-14$\,km s$^{-1}$, the offset is $0.1-0.5$\,dex. Because of the systematic trend in this comparison and the large potential systematic offset, we opt to use the TRES $\log g$ determination in the main text even though it does not employ a correction for macroturbulent broadening (see Section \ref{section:rv}; Table \ref{table:xirt}). Finally, we note that a bias in the APOGEE $\log g$ value could affect other spectroscopically determined parameters for J05215658 discussed in Section \ref{section:cn}.

{\bf Other Analyses:} We note that the analysis by Ref \cite{Ness} finds a higher value for the effective temperature of $T_{\rm eff}\simeq4645.7$\,K,  similar $[{\rm Fe/H}] \simeq-0.311$ with $[{\rm \alpha/Fe}]\simeq 0.159$, and a lower value of  $\log g \simeq2.220$ than APOGEE ASPCAP. The low value of $\log g$ may have been inferred from the fact that J05215658 has a lower $T_{\rm eff}$ than the majority of giants at the nominal APOGEE $\log g$ (see Section \ref{section:tracks}) and a near-Solar value of $[{\rm C/N}]$ (Section \ref{section:cn}). Indeed, the analysis of Ref \cite{Ness} yields a very low mass of  $\ln(M/{\rm M_\odot})\simeq -0.6671$ ($\simeq0.51$\,M$_\odot$) and an unphysical age of $\ln({\rm Age/Gyr})\simeq 4.331$ (76\,Gyr). These values are inconsistent with our determinations of the giant radius and luminosity from the Gaia distance, the argument from $v \sin i$ and tidal synchronization, and the measured proper motions (see Section \ref{section:gaia}).

Ref.\ \cite{Sanders2018} use Gaia and APOGEE, among other spectroscopic surveys, and a Bayesian framework to characterize the probability density  of distance, mass, and age for giants throughout the Galaxy. For J05215658, they find a distance of $1.465$\,kpc (parallax $0.683$) and a mass of $1.24$\,M$_\odot$. Their quoted distance is inconsistent with our 2-$\sigma$ upper limits on the parallax for any value of $\sin i$ in Table \ref{table:parallax}. As we show in Section \ref{section:ruwe}, and in Figures \ref{figure:ruwe} and \ref{figure:trueplx}, the low noise of the Gaia single-star astrometric solution essentially rules out a true parallax as large as $0.683$.

\begin{figure*}
\centerline{\includegraphics[width=8.cm]{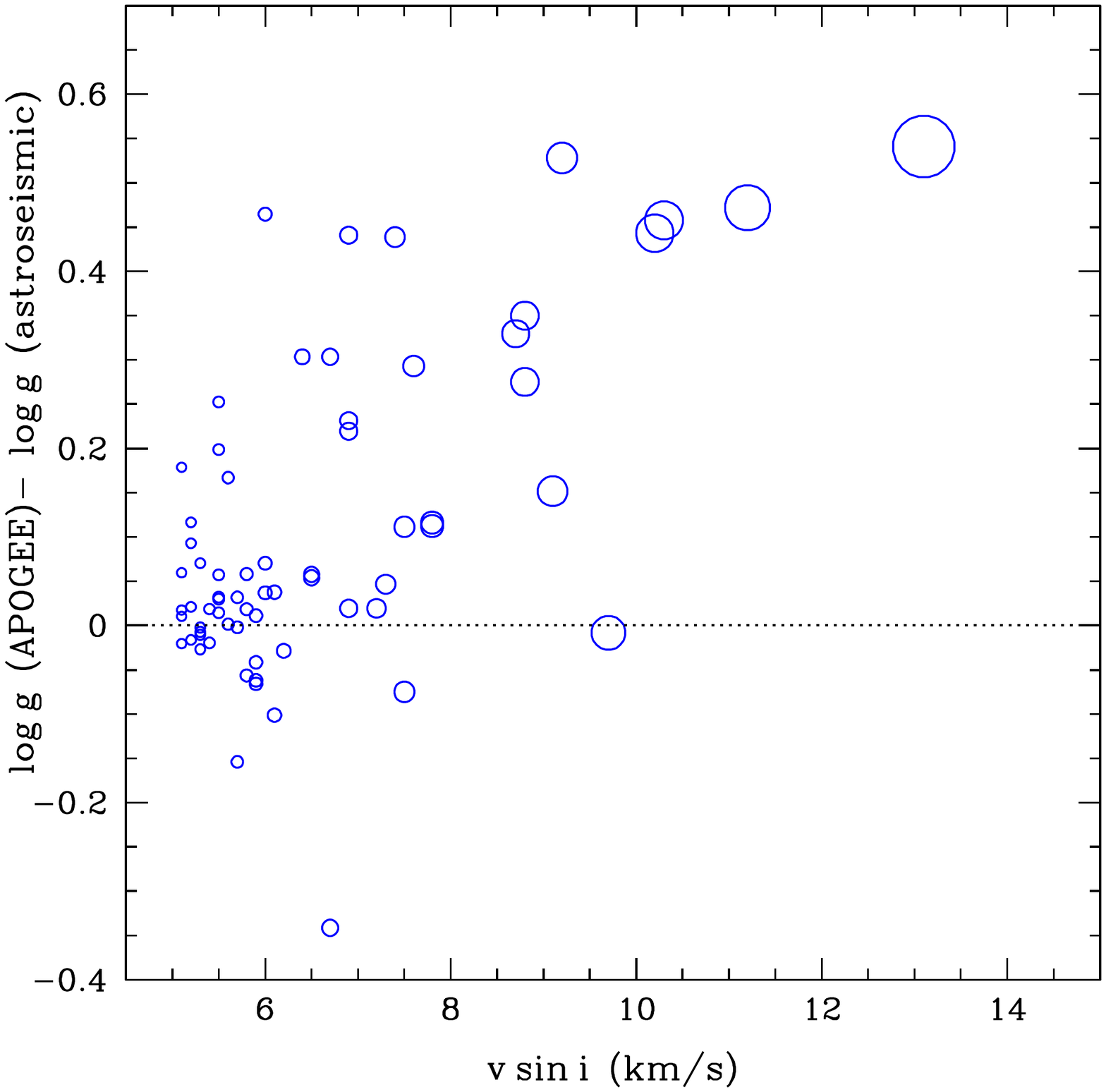}\includegraphics[width=8.cm]{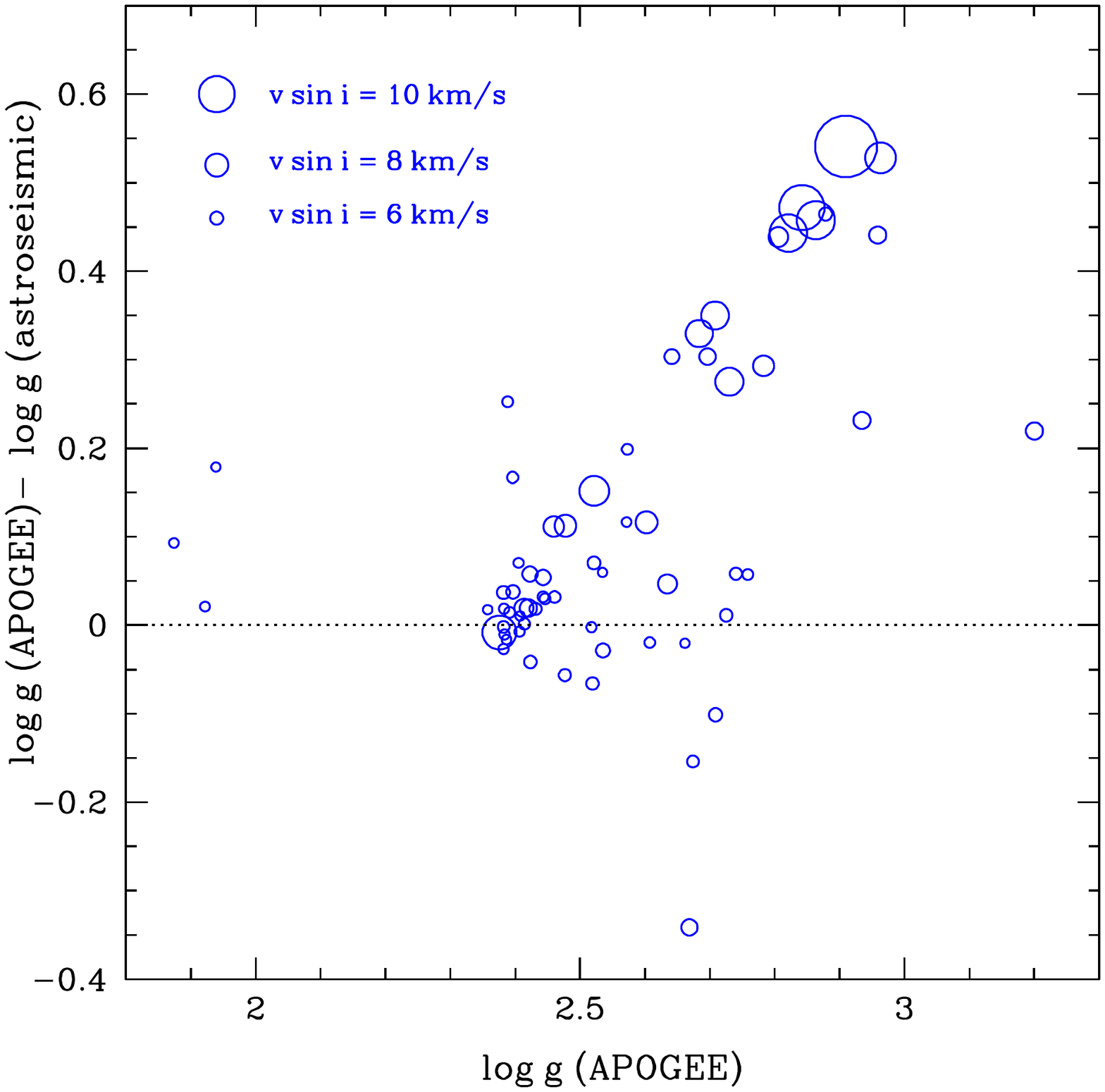}}
\vspace*{-2.5cm}
\centerline{\includegraphics[width=8.cm]{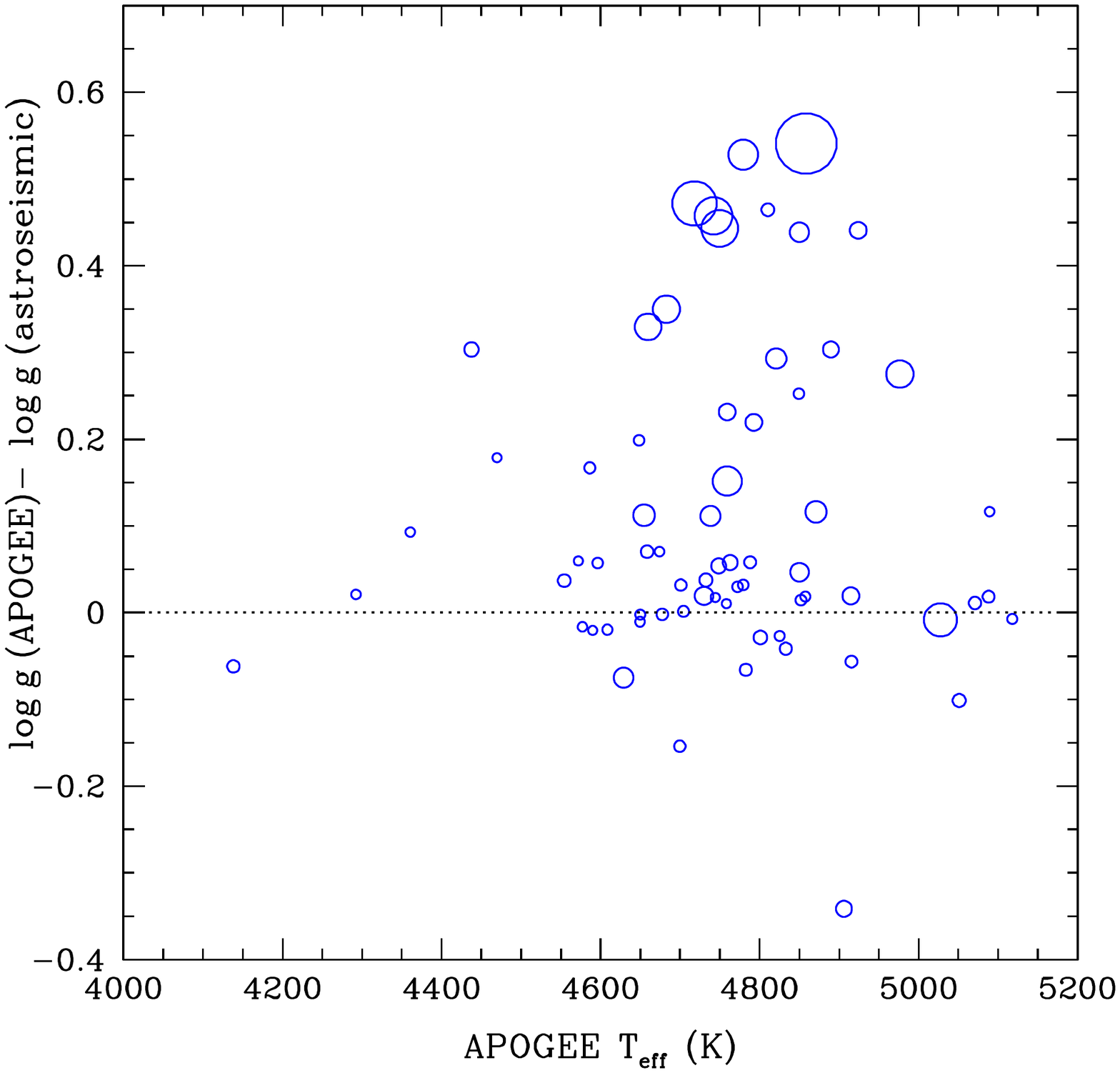}\includegraphics[width=8.cm]{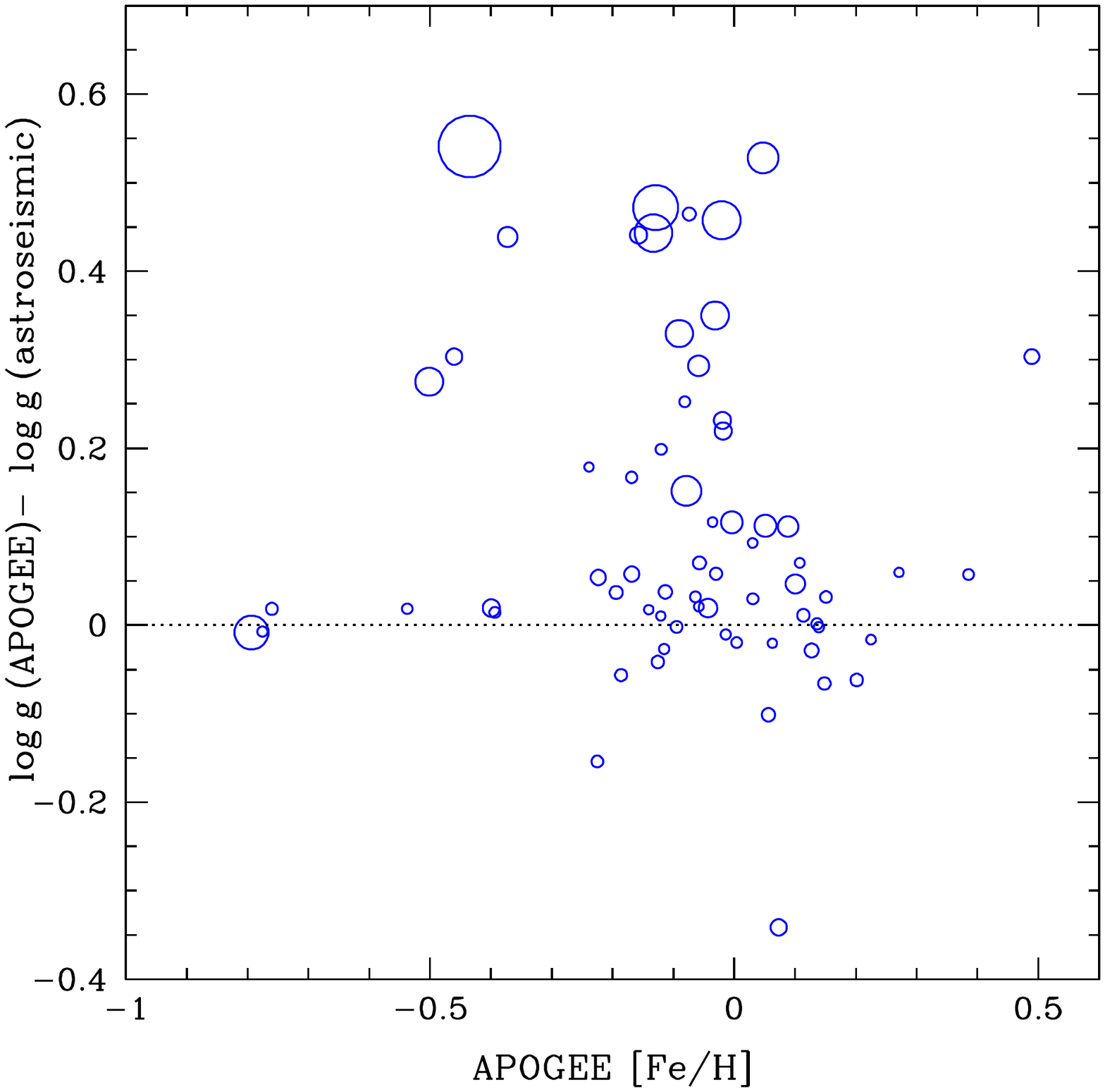}}
\vspace*{-1.5cm}
\caption{All panels show the difference between the APOGEE DR14 $\log g$ and the $\log g$ as measured by astroseismology as a function of $v \sin i$ from the APOKASC sample \cite{Pinsonneault18}, matched to those stars with $v \sin i$ measurements from \cite{Tayar}. In each panel the size of the circle is scaled by the the value of $v \sin i$. The top left, top right, bottom left, and bottom right panels show this difference in $\log g$ as a function of $v \sin i$, APOGEE $\log g$, $T_{\rm eff}$, and $[{\rm Fe/H}]$, respectively. These plots demonstrate that APOGEE systematically overestimates $\log g$ for rapidly rotating stars in the range of parameters appropriate for J05215658 ($v \sin i \simeq 14$\,km/s, $T_{\rm eff}\simeq4500$\,K, $[{\rm Fe/H}]\simeq-0.4$, $\log g$ (APOGEE)$\simeq2.6$). The upper left panel shows that there is a systematic trend as a function of $v \sin i$.}
\label{figure:rot}
\end{figure*}

\subsection{Abundances}
\label{section:cn}

The derived abundances from APOGEE DR13 are shown in Table \ref{table:abundances}. The system is observed to be metal poor, but with near-Solar $[{\rm C/N}]\simeq0.034$, and modest enhancements in S and O with respect to Fe of $[{\rm S/Fe}]\simeq0.244$ and $[{\rm O/Fe}]\simeq0.118$.

Figure \ref{figure:cn} shows the $[{\rm C/N}]$ abundances as a function of astroseismic mass from the APOKASC sample \cite{Pinsonneault18}. As discussed in the main text, the fact that J05215658 has $[{\rm C/N}]\simeq0.0$ implies a low $M_{\rm giant}\simeq1$\,M$_\odot$ in the absence of other information. Note that the bias in the APOGEE $\log g$ discussed in Section \ref{section:apogee} (Fig.~\ref{figure:rot}) may affect the abundance determinations and other spectroscopic parameters.

Three stars in the APOKASC sample have astroseismic $M_{\rm giant}> 2.0$\,M$_\odot$ {\it and} $[{\rm C/N}]\ge-0.1$ (KIC 8649099, 11954055, and 9541892 with masses of 2.0, 2.7, and 3.1\,M$_\odot$, respectively). Although such objects are well away from the mean trend in $[{\rm C/N}]$ versus mass, a significant fraction of the massive stars in the sample also have high $[{\rm C/N}]$. Such stars may be the result of stellar mergers \cite{Izzard}. The fraction of stars in the sample with high values of $[{\rm C/N}]$ apparently increases as a function of the giant mass:  while only $8/1577\simeq0.005$ of all $>1.5$\,M$_\odot$ giants in APOKASC have $[{\rm C/N}]>-0.1$ the fraction increases to $2/135\sim0.015$ for $M_{\rm giant}>2.5$\,M$_\odot$ and to $1/18\sim0.06$ for $M_{\rm giant}>3.0$\,M$_\odot$. The shaded regions in Figure \ref{figure:cn} denote the 1- and 2-$\sigma$ confidence intervals of $M_{\rm giant}$ from Figure \ref{figure:m}.

\begin{table}[!t]
\begin{center}
\caption{APOGEE DR13 Abundances for J05215658. See discussion in Sections \ref{section:apogee} and  \ref{section:cn}.
\label{table:abundances}}
\begin{tabular}{lccc}
Element & Abundance & Uncertainty  \\
\\
\hline
\hline
\\
$[{\rm Al/H}]$ & $-0.769$ & $ 0.069 $ \\
$[{\rm Ca/H}]$ & $ -0.457$ & $  0.034$  \\
$[{\rm C/H}]$ & $ -0.321 $ & $ 0.041$  \\
$[{\rm Fe/H}]$ & $ -0.503$ & $  0.036$  \\
$[{\rm K/H}]$ & $ -0.629 $ & $ 0.059 $ \\
$[{\rm Mg/H}]$ & $ -0.428 $ & $ 0.028$  \\
$[{\rm Mn/H}]$ & $ -0.355 $ & $ 0.041$  \\
$[{\rm Na/H}]$ & $ -0.518$ & $  0.102 $ \\
$[{\rm Ni/H}]$ & $ -0.402$ & $  0.042$  \\
$[{\rm N/H}]$ & $ -0.355 $ & $ 0.074$  \\
$[{\rm O/H}]$ & $ -0.385 $ & $ 0.027$  \\
$[{\rm Si/H}]$ & $ -0.605$ & $  0.040$  \\
$[{\rm S/H}]$ & $ -0.259 $ & $ 0.042 $ \\
$[{\rm Ti/H}]$ & $ -0.633$ & $  0.048$  \\
$[{\rm V/H}]$ & $ -0.632$ & $  0.135$  \\
\\
\hline
\hline
\end{tabular}
\end{center}
\end{table}

\subsection{Comparison with Single-Star Theoretical Evolutionary Tracks}
\label{section:tracks}

Figure \ref{figure:mist} shows $L$ (left) and $\log g$ (right) versus $T_{\rm eff}$ for stellar evolutionary models of different masses and for $[{\rm Fe/H}]=-0.4$ (top) and 0.0 (bottom). The shaded regions indicate $T_{\rm eff}\simeq4525.0\pm90$\,K, $\log g=2.35\pm0.14$, and $L=331^{+227}_{-125}$\,L$_\odot$, as inferred from TRES (Section \ref{section:tres_star}), the  Gaia parallax (Section \ref{section:gaia}), and the bolometric flux and SED (Section \ref{section:sed}). The low value of the effective temperature may be interpreted as favoring a lower mass giant of $\sim1$\,M$_\odot$ (right panels), whereas the bolometric luminosity strongly favors a higher mass giant of $\sim2-3$\,M$_\odot$ when comparing by-eye to the MIST tracks. 

Quantitative fits to single-star evolutionary tracks are discussed in the main text. Given the nominal values and uncertainties in $L$, $T_{\rm eff}$, and $R$ from the Gaia parallax and our fit to the SED, and using the TRES value of $\log g\simeq2.35\pm0.14$ as a constraint, we find a best fit of $M_{\rm giant}\simeq3.2^{+1.0}_{-1.0}$\,M$_\odot$ (2-$\sigma$). Using the fit value for the giant radius $R$, and comparing with the minimum radius obtained from $v\sin i$, we derive a constraint on $\sin i$ that allows us to constrain the companion mass to be $M_{\rm CO}\simeq3.3^{+2.8}_{-0.7}$\,M$_\odot$. For a given assumed value of $v\sin i$, some fraction of the fitted evolutionary models have an unphysical $\sin i>1$. As discussed in the main text, if we assume a value of the macroturbulent broadening large enough to give a $v \sin i=10$\,km s$^{-1}$, corresponding to $\xi_{\rm RT}>10$\,km s$^{-1}$ in Table \ref{table:xirt_apogee}, more low-mass giant models have physical values of  $\sin i<1$, and best-fit giant masses of $\simeq1.8$\,M$_\odot$ can be obtained. However, these lower-mass giant models do not have much lower companion masses because the lower implied value of $\sin i$ drives up the companion mass $M_{\rm CO}$, in accord with the mass function (Fig.\ \ref{figure:m}). For the models we have explored, this leads to $M_{\rm CO}>2.5$\,M$_\odot$.

In addition, our results do not change qualitatively if we change the constraint on $\log g$, or impose no constraint on $\log g$ at all. In both cases, we find that the best fits for the giant mass decrease to the lower end of the range quoted when imposing the TRES $\log g$, with best-fit giant masses in the range of 
$M_{\rm giant}\simeq2.2-2.5$\,M$_\odot$. Imposing $v \sin i$, these fits then give $\sin i\simeq0.8-0.9$ and best-fits to the compact object companion mass of $M_{\rm CO}\simeq2.9-4.0$\,M$_\odot$.

\subsection{Limits on the Giant Radius from Ellipsoidal Variations}
\label{section:ellipsoidal}

We do not convincingly detect ellipsoidal modulations in the ASAS-SN lightcurve. Using a periodogram search and Lomb-Scargle analysis, the ASAS-SN lightcurve exhibits a small peak in power at a period of $\simeq83.2/2$\,days as expected for ellipsoidal variations, but when we subtract the dominant periodicity associated with the spot modulation, we find a phase for the modulation that is inconsistent with ellipsoidal variations: the maximum photometric variation is different from the maximum RV blueshift by $\simeq30$\,degrees. To assess whether or not we could in fact see ellipsoidal variations of a given amplitude, we injected periodic modulations into the ASAS-SN photometry (Figs.\ \ref{figure:lc} and \ref{figure:raw}) consistent with the phase and period of the radial velocity curve for ellipsoidal variations of specified V-band amplitude, using the cadence and photometric errors from the actual ASAS-SN observations. Then, using the same types of searches, we look for power with the specified period and phase. Using these explorations, we find that the signal at $\simeq83.2/2$\,days in the ASAS-SN lightcurve has peak-to-peak amplitude of order $\simeq3$\%, and that this is close to the minimum we could detect.

With this upper bound on the photometric variations for which we would expect to see evidence of periodic modulations consistent with ellipsoidal variability, we can construct a  bound on the giant radius using \cite{Morris1985}
\beq
R^3=\frac{3.070\,A^3\,\Delta M\,(3-u)}{q\sin^2 i\,(\tau+1)(15+u)},
\eeq
where $A$ is the semi-major axis in Solar radii, $q=M_{\rm CO}/M_{\rm giant}$, $\Delta M$ is the peak-to-peak variation in the lightcurve, $u\simeq0.83$ is the limb-darkening coefficient, $\tau\simeq0.46$ is the gravity-darkening coefficient assuming a late-type star with a convection envelope, $T_{\rm eff}=4500$\,K, and V-band observations. For $\Delta M\lesssim0.03$, we find that $R\lesssim30$\,R$_{\odot}$ for $\sin i=1$ and $q=1$. 

In our fits to the evolutionary tracks described in the main text, we find that the giant radii derived over our preferred parameter regime are always small enough that they can accommodate the peak-to-peak variability limit of $\simeq3$\% we infer from the ASAS-SN photometry. However, these fits in general produce stellar radii, component masses, and semi-major axes predicting that the ellipsoidal variability should appear at the $\sim1$\% level.

\subsection{Spectral Energy Distribution \& UV Detection}
\label{section:sed}

Figure \ref{figure:sed} shows the spectral energy distribution (SED). We fit the WISE 3.4 and 4.6\,$\mu$m  \cite{WISE}, 2MASS J, H and K$_s$ \cite{2MASS}, our BV$ri$ and the U-band Swift photometry using model atmospheres with $[{\rm Z/H}]=-0.5$  and $\log g = 2.5$  \cite{Castelli2004}, assuming 10\% flux errors to compensate for variability, the Gaia distance of $3.11$\,kpc (parallax 0.322\,mas; eq.\ \ref{best_parallax}), an $R_V=3.1$ extinction law \cite{Cardelli1989}, and a spectroscopic temperature of $T_{\rm eff}=4550 \pm 100$\,K (Section \ref{section:giant}).   This process yielded a luminosity of $\log(L/L_\odot)=2.52 \pm 0.03$, a temperature of $T_{\rm eff}=4530 \pm 89$\,K, and an extinction of $E(B-V)=0.445 \pm 0.050$. Note that the photometry has slightly improved the constraint on the temperature, and that the extinction is consistent with estimates for this distance from the three dimensional Pan-STARRS1 dust maps  \cite{Green2015}. The data points and solid red line show the spectral energy distribution of the giant and our best fit at the nominal {\it Gaia} distance. The goodness of fit is $\chi^2=5.83$ for 8 degrees of freedom. The excellent fit to the SED at the spectroscopic temperature and including significant Galactic extinction with the standard $R_V=3.1$ extinction law also implies that the dust properties in the direction of J05215658 cannot be unusual. In addition, the WISE $12$ and $22\,\mu$m fluxes (not shown) lie on the red extension of the SED model so there is no infrared excess indicating the presence of significant circumstellar dust and extinction. The bolometric flux used throughout the text is $F\simeq1.1\times10^{-9}$\,ergs cm$^{-2}$ s$^{-1}$.

Figure \ref{figure:sed} shows the best-fit resulting SED as the solid red curve at the Gaia distance of $3.11$\,kpc. The blue dashed lines show SEDs for main sequence companions of 1.3, 1.4,1.5, and 1.8\,M$_\odot$ at an age of 1\,Gyr for comparison. The dotted lines show the sum of the main sequence models and the best-fitting SED to demonstrate how a main sequence companion would effect the bluer bands at the Gaia distance. If the parallax bias induced by the binary orbital motion is negative (see Section \ref{section:gaia}), and the true parallax of the system is larger, then the fitted luminosity of the giant decreases and the spectral distortions to the bluer bands caused by assuming a main sequence companion increase (see Section \ref{section:limits}). No infrared excess is apparent.

{\bf UV Detection:} Note that the Swift UVM2 detection was not included in the SED fit. The mean magnitude from our observations is reported in Table \ref{table:giant_photometry} and shown for the Gaia distance in Figure \ref{figure:sed}, but we were unable to find a satisfactory fit to the SED template when we tried to include it. GALEX reports an NUV detection at $\simeq21.5\pm0.4$\,mag, which we also include in Figure \ref{figure:sed}, falling approximately a factor of $\sim3$ below our mean Swift UVM2 flux. In addition, we find evidence for variability in the UVM2 band from our multi-epoch Swift follow-up observations. Table \ref{table:uvm2} gives the date of the Swift observation, derived magnitude or upper limit, and uncertainty in each observation. While our first observation gave a $>20.27$\,mag 3-$\sigma$ upper limit, subsequent observations yield detections of $\simeq20.2 - 19.8$\,mag.   The brightness variation is inconsistent with the known multi-band variability from Figure \ref{figure:post} (top panel) in both amplitude and phase. We ran photometry for two other stars in the field during our observations to check if the implied UVM2 variability of J05215658 might be an artifact of the observations or image processing.  One star showed virtually constant flux over all observations, while the other showed some variability, but opposite the implied trends derived for J05215658: while the test star became brighter from one epoch to the next, J05215658 became dimmer.

We thus conclude that J05215658 is variable in the UVM2 Swift band. A number of interpretations can be considered. In Section \ref{section:limits} we consider the possibility of a stellar companion, and show that main sequence or stripped envelope stars cannot simultaneously explain the UV photometry and the RV curve. Wind-fed accretion onto a neutron star or black hole could also be considered, but the X-ray upper limit described in Section \ref{section:x} constrains this possibility.  The simplest interpretation of the variable UVM2 detection is that J05215658 has some level of activity, which is common for rapidly rotating giants. In particular, Ref.\ \cite{Bai} shows that giants with $T_{\rm eff}$ and $\log g$ similar to J05215658 commonly exhibit both UV excesses and variability.

\begin{table}[!t]
\begin{center}
\caption{Multi-epoch Swift UVM2 Photometry (see Section \ref{section:sed}; Table \ref{table:giant_photometry}).
\label{table:uvm2}}
\begin{tabular}{lccc}
Observation Date & UVM2 & uncertainty \\
(MJD) & (mag) & (mag)\\
\\
\hline
\hline
\\
58083.047  &  $>20.27$  &  \\
58339.351  &  $19.75$  & $0.21$\\
58340.814  &  $20.22$  & $0.21$\\
58343.001  &  $19.90$  & $0.19$\\
58345.067  &  $20.06$  & $0.22$\\
58369.037  &  $19.77$  & $0.18$\\
\\
\hline
\hline
\end{tabular}
\end{center}
\end{table}

\subsection{Limits on a Stellar Companion}
\label{section:limits}

As shown in Figure \ref{figure:m}, for $\sin i=1$ and $M_{\rm giant}=1$\,M$_\odot$, the minimum mass of the unseen companion allowed from the radial velocity measurements is $\simeq1.8$\,M$_\odot$. The blue lines in Figure \ref{figure:sed} show the spectral energy distribution of $1.3$, $1.4$, $1.5$, and $1.8$\,M$_\odot$ main sequence stars at 1 Gyr \cite{Castelli2004} compared to our fit to the photometry at the nominal {\it Gaia} distance of $3.11$\,kpc. 

The Swift UVM2 detection is very constraining (Table \ref{table:giant_photometry}). Main sequence stellar companions of $>1.4$\,M$_\odot$ are ruled out. While a lower mass $<1.4$\,M$_\odot$ companion is nominally consistent with the UVM2 limit, this mass is inconsistent with the results from Figure \ref{figure:m} unless the giant mass is $M_{\rm giant}\simeq0.2$\,M$_\odot$ ($0.5$\,M$_\odot$) for $\sin i=0.9$ ($\sin i=1.0$). Such a low value for the giant mass is implausible given the distance to the system, the luminosity of the giant, and its evolutionary state.

As discussed in Section \ref{section:gaia} there may be a bias in the Gaia parallax as a result of astrometric binary orbital motion (see eq.~\ref{best_parallax}). If the distance is in fact smaller than the nominal value of 3.1\,kpc used for Figure \ref{figure:sed}, then the photometric limits on a main sequence companion become tighter because the data points and SED model would move to lower $\lambda L_\lambda$, while the main sequence stellar models would remain unchanged. For example, if the distance to the system was $D\simeq2.0$\,kpc instead of $3.1$\,kpc, the $1.4$\,M$_\odot$ main sequence companion would be excluded and the $1.3$\,M$_\odot$ model would be at the mean Swift UVM2 detection. However, the U, B, and V bands would then be poorly fit, and it would be impossible to explain the radial velocity variation. Indeed, we can only accommodate a higher-mass companion photometrically if the distance is {\it underestimated} by Gaia.  For example, for a $1.8$\,M$_\odot$ main sequence companion star, which would satisfy the dynamical constraints from the RV measurements if $M_{\rm giant}=1$\,M$_\odot$ ($\sin i=1$), to be consistent with the photometry, the distance to the system would need to be $>2$ times larger than $3.1$\,kpc, and the luminosity of the giant would then need to be more than 4 times larger, over $10^3$\,L$_\odot$. Such a luminosity would then be inconsistent with a giant mass as low as the $M_{\rm giant}=1$\,M$_\odot$ needed to accommodate a companion of $1.8$\,M$_\odot$. We therefore see no way to have a main sequence companion that satisfies the dynamical, photometric, and astrometric constraints on the system. As discussed in Section \ref{section:sed}, the UVM2 detection, and its variability, is more reasonably interpreted as intrinsic variability of the rapidly rotating giant star.

Much cooler stellar companions that evade the Swift UVM2 detection rapidly degrade the fit to the SED unless they have the same temperature as the giant. In that case, the companion would also have to be a giant star, in which case we would expect it to show up in the TRES and APOGEE spectroscopy. However, neither the TRES nor the APOGEE spectra (Section \ref{section:rv}) show any evidence for a second set of spectral lines at any of the radial velocity epochs (Figure \ref{figure:post}; Table \ref{table:rv}). Moreover, in all of the allowed parameter space of Figure \ref{figure:m} for $M_{\rm giant}<3$\,M$_\odot$, the unseen putative red giant companion would have to be more massive than the observed giant. Finally we see no evidence of an excess in the NIR SED that might be evidence of a massive cooler companion.

One can also consider much hotter companions. We considered whether stripped stellar cores might be able to meet the dynamical and photometric constraints. For example, in Ref.\ \cite{gotberg} we see that a $1.8$\,M$_\odot$ stripped core has a bolometric luminosity of $\sim10^{3}-10^{3.5}$\,L$_\odot$ and an effective temperature of $50-60$\,kK. Such a model would exceed the Swift UVM2 detection in Figure \ref{figure:sed} by a factor of $\sim100$, dominate the U-band flux, and contribute to the bluer optical bands.  At fixed luminosity of $\sim10^{3}-10^{3.5}$\,L$_\odot$, the effective temperature of such a stripped core would have to be $>4-5$ times higher to accommodate the Swift UVM2 detection. In addition, for the high luminosities expected for a $1.8$\,M$_\odot$ core, we may expect to see bright optical emission lines associated with the strong ionizing flux, which are not present in the observed spectra. We conclude that it is not possible to satisfy both the dynamical and photometric constraints with a stripped very hot stellar core.

\subsection{X-Ray Upper Limit \& Wind-Fed Accretion}
\label{section:x}

For an interacting binary, we would expect ongoing X-ray emission from accretion. While there are weak limits on the X-ray emission from J05215658 from the \textit{ROSAT} All-Sky Survey (RASS) \cite{1999A&A...349..389V}, we obtained much stronger limits from the X-ray Telescope (XRT; \cite{2004SPIE.5165..217H, 2005SSRv..120..165B}) on board the \textit{Neil Gehrels Swift Gamma-ray Burst Mission} observatory \cite{2004ApJ...611.1005G} made simultaneously with the UVOT observations discussed in Section \ref{section:limits}. This observation (ObsID:00010442001), taken on 2017-11-26 01:07:21 UTC (${\rm MJD=58083.05}$), was reprocessed from level one XRT data using the \emph{Swift} \textsc{xrtpipeline} version 0.13.2 script, and with the most up to date calibration files, following standard filter and screening criteria suggested by the \textit{Swift} collaboration.\footnote{\textit{Swift} XRT data reduction guide: \url{http://swift.gsfc.nasa.gov/analysis/xrt_swguide_v1_2.pdf}}.

We find no evidence for X-ray emission associated with J05215658 to the upper limit reported in Table \ref{table:giant_photometry}. There is a faint nearby X-ray point source located at $(\alpha,\delta)=(05^{h}21^{m}56.6^{s}, +43^{\circ}49'22'')$, approximately $30$\,arcsec away from the system. To minimize contamination from this nearby source when deriving our $3-\sigma$ count-rate upper limit, we use a source region with a radius of 20 arcsec centered on J05215658 and a source free-background region centered at  $(\alpha,\delta)=(05^{h}22^{m}30.9^{s},$ $+43^{\circ}56'40.3'')$ with a radius of 150 arcsec. Correcting for the fraction\footnote{A 20 arcsec radius corresponds to an encircled energy fraction of $\sim$ 80\% at 1.5 keV  assuming on-axis pointing \cite{2004SPIE.5165..232M}.} of the total counts from the system that would be enclosed by our source region, we obtain an upper limit of $5\times10^{-4}$\,counts/sec in the $0.3-10.0$\,keV energy band. Assuming an absorbed powerlaw with a photon index of $\Gamma=2$, and a Galactic H\textsc{i} column density of $4.03\times10^{21}$\,cm$^{-2}$ derived from \cite{2005A&A...440..775K}, this count rate corresponds to an unabsorbed flux limit of $4.4\times10^{-14}$\,ergs s$^{-1}$ cm$^{-2}$, or $\simeq1\times10^{-2}$\,L$_\odot$ at 3.1\,kpc, roughly $10^7$ times smaller than the Eddington luminosity for a 3\,M$_\odot$ black hole.

We considered the possibility of wind-fed accretion from the giant to the compact object companion. We scale the wind mass loss rate to $\dot{M}_{\rm wind,\,-10}=\dot{M}_{\rm wind}/10^{-10}$\,M$_\odot$ yr$^{-1}$, with a wind velocity of $V_{\rm wind,\,200}=V_{\rm wind}/200$\,km s$^{-1}$, approximately the escape velocity for a $3$\,M$_\odot$ star with $R=25$\,R$_\odot$. The total separation between the two bodies we take to be $S=0.68$\,AU$/\sin i$ (eq.\ \ref{motion}). An estimate of the amount of material gathered at the sphere of influence of the compact object is 
\beq
\dot{M}_{\rm acc}\sim \frac{\dot{M}_{\rm wind}}{(4\pi S^2)}\,\pi \left(\frac{G M_{\rm CO}}{V_{\rm wind}^2}\right)^2\sim2\times10^{-13}\,{\rm M_\odot\,\,yr^{-1}}\,\,\,\frac{\dot{M}_{\rm wind,\,-10}\,M_{\rm CO,\,3}^2\sin^2 i}{V_{\rm wind,\,200}^{4}}
\eeq
where $M_{\rm CO,\,3}=M_{\rm CO}/3$\,M$_\odot$. For radiatively efficient accretion onto a black hole, we would expect an accretion luminosity of order $\sim0.1\dot{M}_{\rm acc}c^2 \sim 0.35$\,L$_\odot$. Although this is above the X-ray upper limit, for such low accretion rates the flow will be radiatively inefficient \cite{Narayan}. For a massive neutron star companion with even a small surface dipole magnetic field, the energy density of the field greatly exceeds ram pressure of the accreted material at the neutron star surface, implying that much of the gas may be expelled from the system without accreting.

\begin{figure*}
\centerline{\includegraphics[width=8.5cm]{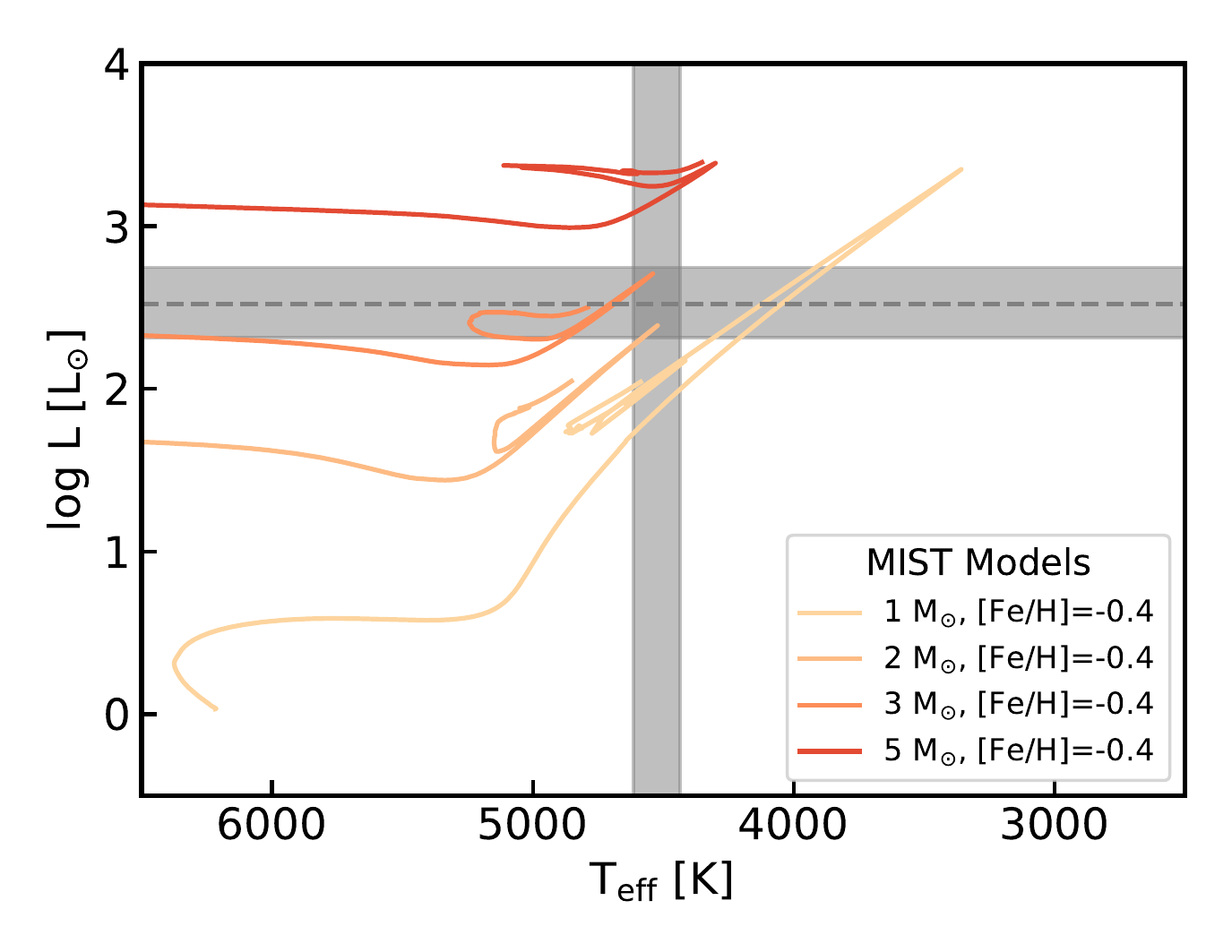}\includegraphics[width=8.5cm]{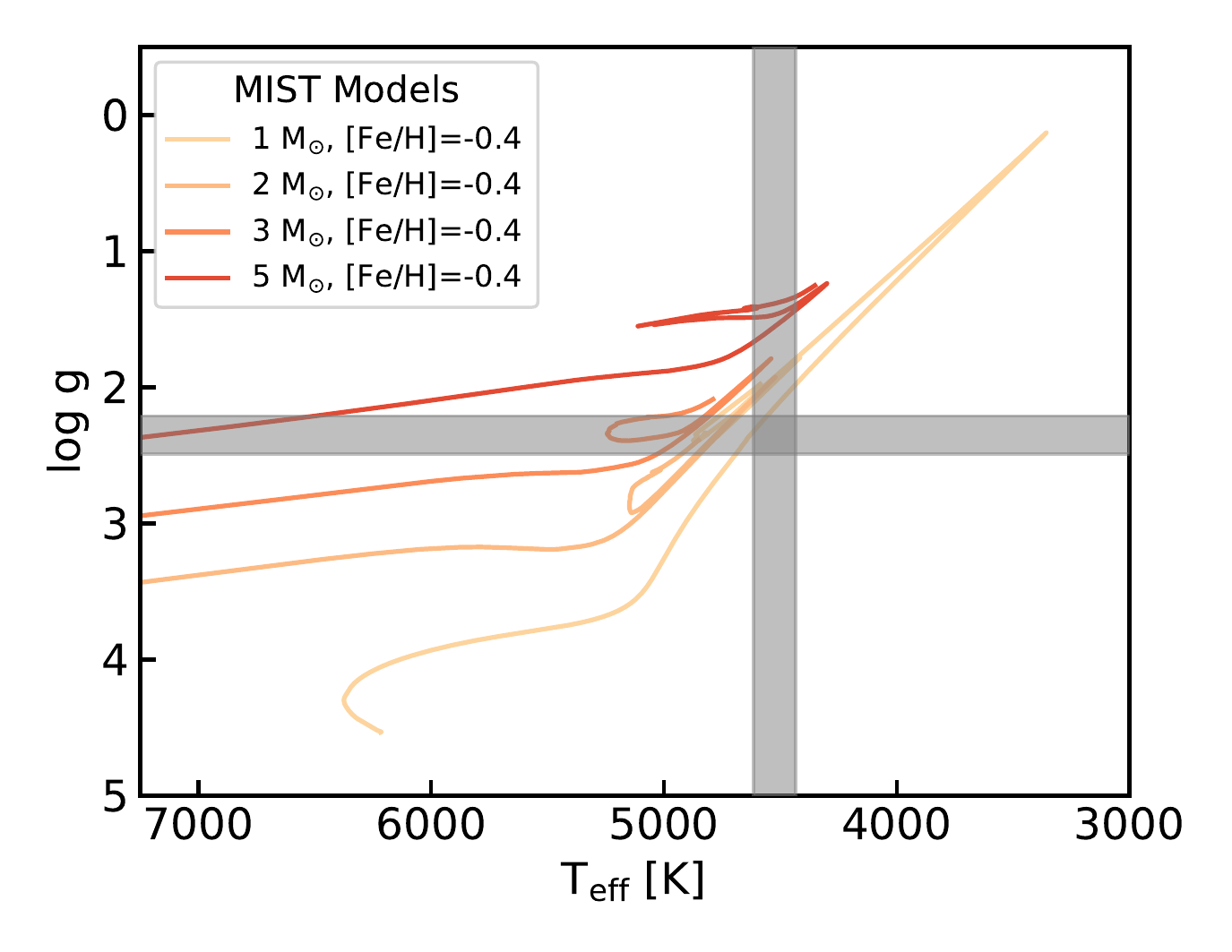}}
\centerline{\includegraphics[width=8.5cm]{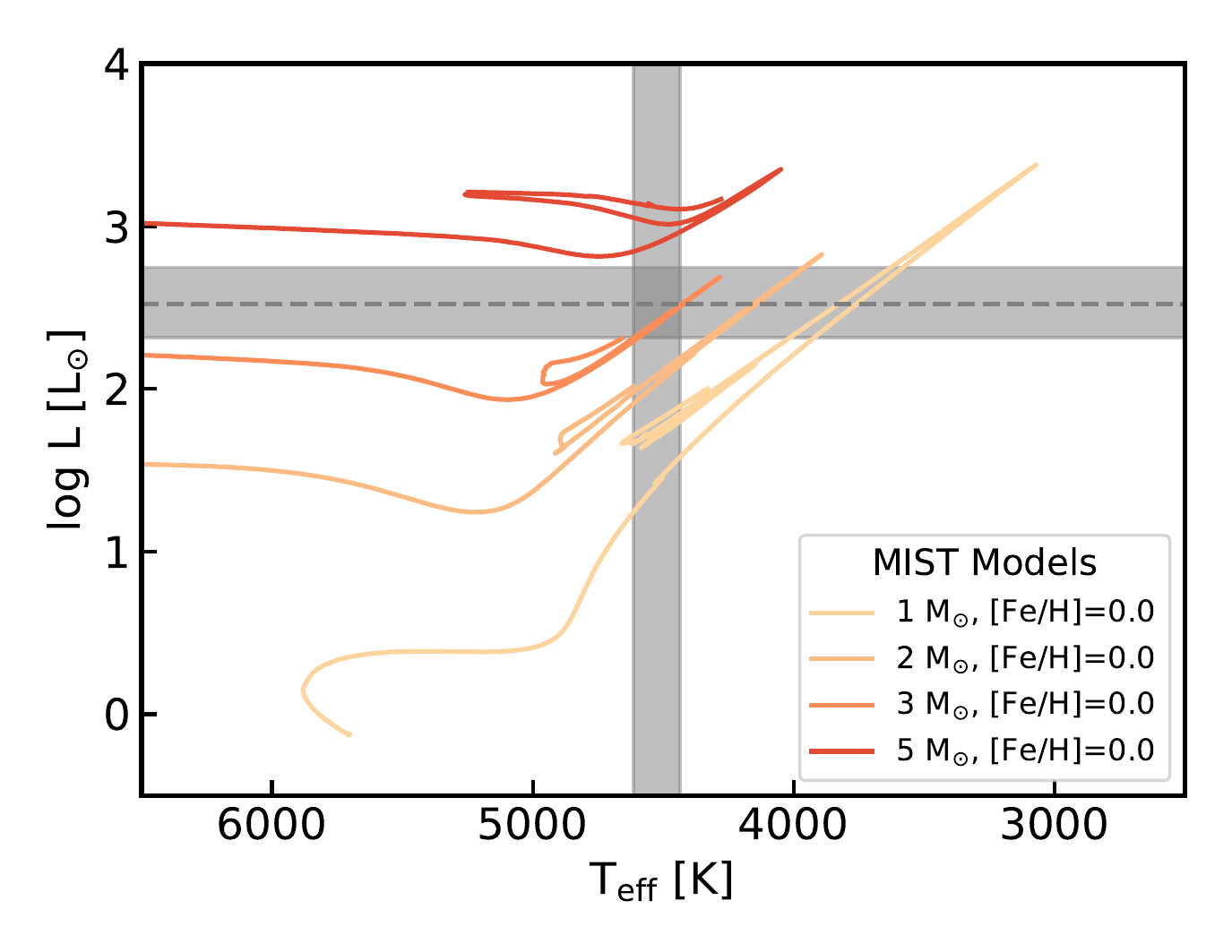}\includegraphics[width=8.5cm]{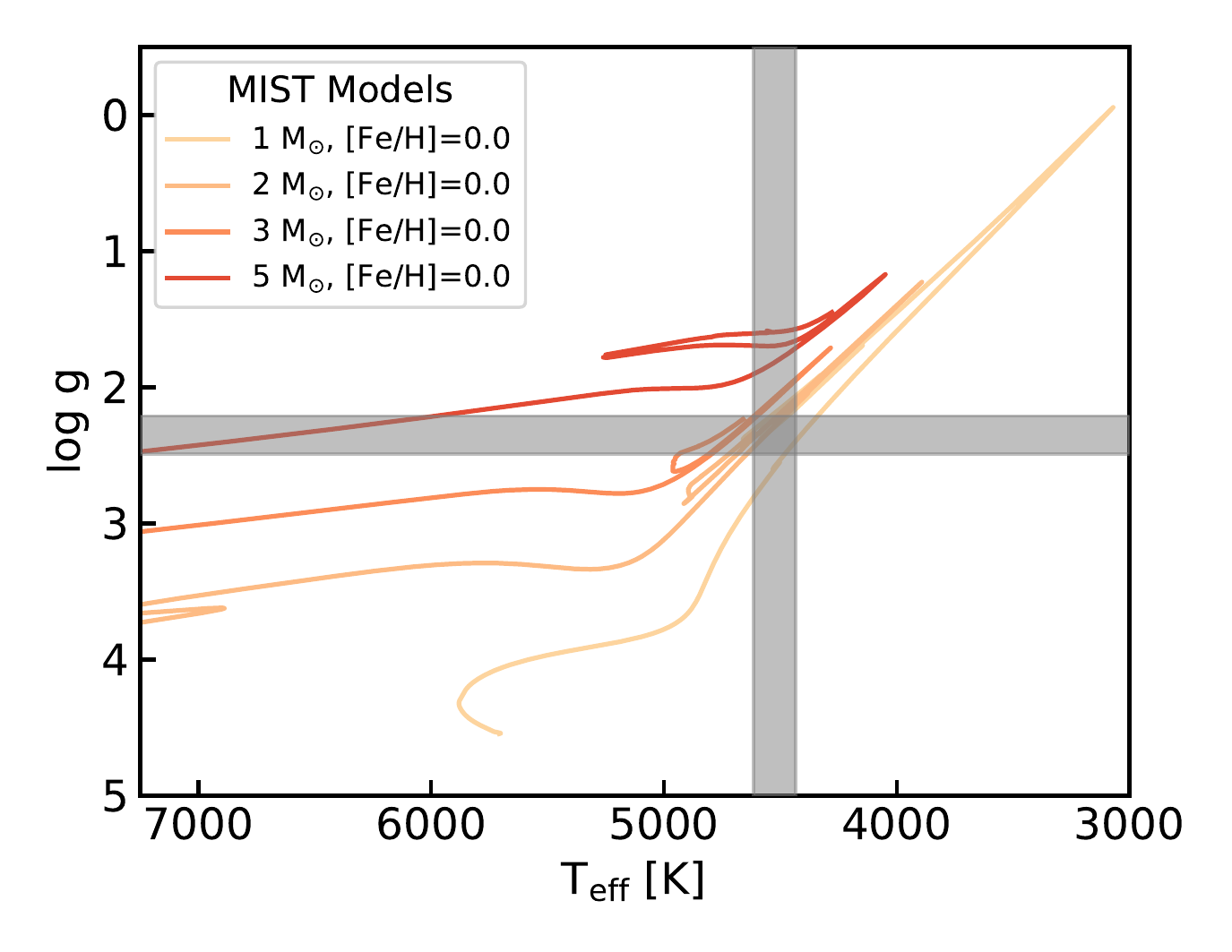}}
\caption{Bolometric luminosity (left) and $\log g$ (right) as a function of $T_{\rm eff}$ for MIST single-star models with $[{\rm Fe/H}]=-0.4$ (top panels) and $[{\rm Fe/H}]=0$ (bottom panels) for a range of masses from $1-5$\,M$_\odot$  \cite{Paxton1,Paxton2,Dotter,Choi}. Grey bands indicate properties of the giant with $T_{\rm eff}=4525\pm90$\,K and $\log g=2.35\pm0.14$ (Section \ref{section:tres_star}). The horizontal dashed line and gray band indicate the bolometric luminosity $L\simeq331^{+227}_{-125}$\,L$_\odot$ of the giant as inferred in the main text from the {\it Gaia} distance and observed bolometric flux. See discussion in main text and Section \ref{section:apogee}.}
\label{figure:mist}
\end{figure*}

\begin{figure*}
\centerline{\includegraphics[width=14.cm]{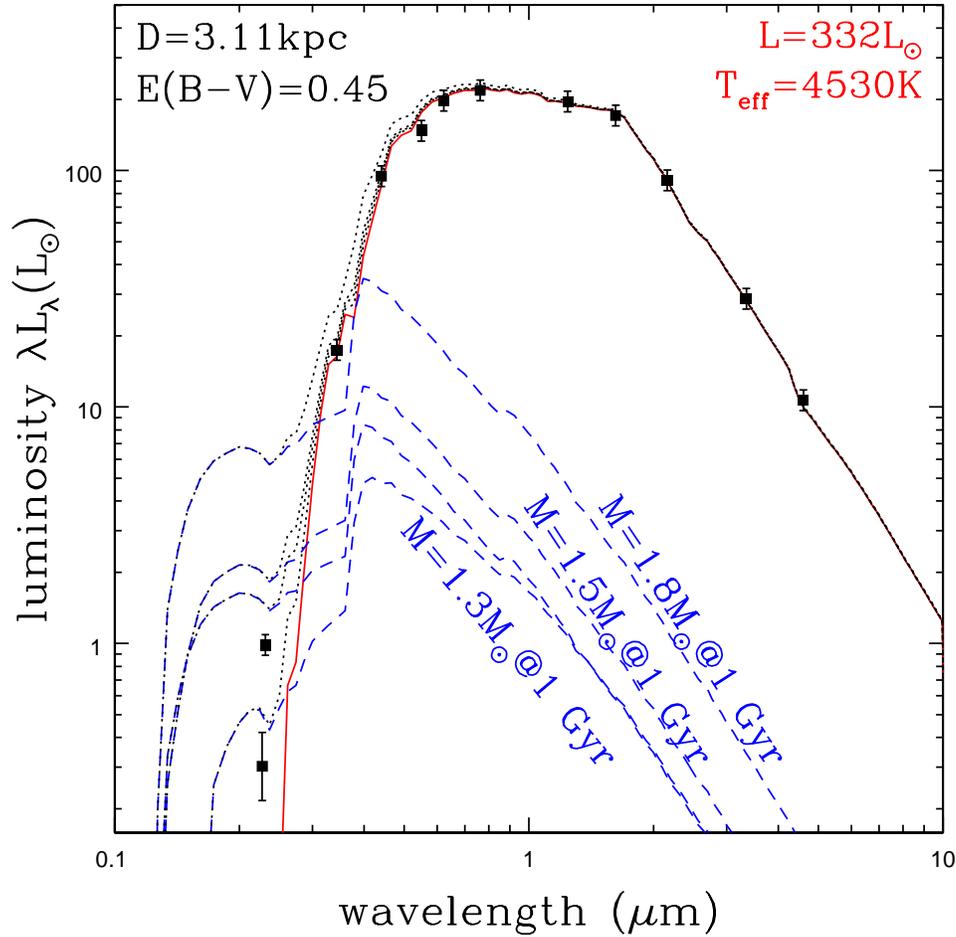}}
\vspace*{-2.5cm}
\caption{Spectral energy distribution (SED) of J05215658 normalized for the nominal {\it Gaia} distance of 3.11\,kpc (eq.~\ref{parallax}; data points) with a fit to the SED (red solid line) as described in Section \ref{section:sed} with fit parameters labelled. The blue dashed lines show SEDs for main sequence companions of 1.3, 1.4, 1.5, and 1.8\,M$_\odot$ for comparison.  The dotted black lines show the sum of the best-fitting SED and the main sequence models. Reddened giant templates cannot accommodate the UVM2 Swift detection. See Sections \ref{section:sed} and \ref{section:limits}.}
\label{figure:sed}
\end{figure*}

\begin{figure*}
\centerline{\includegraphics[width=14.cm]{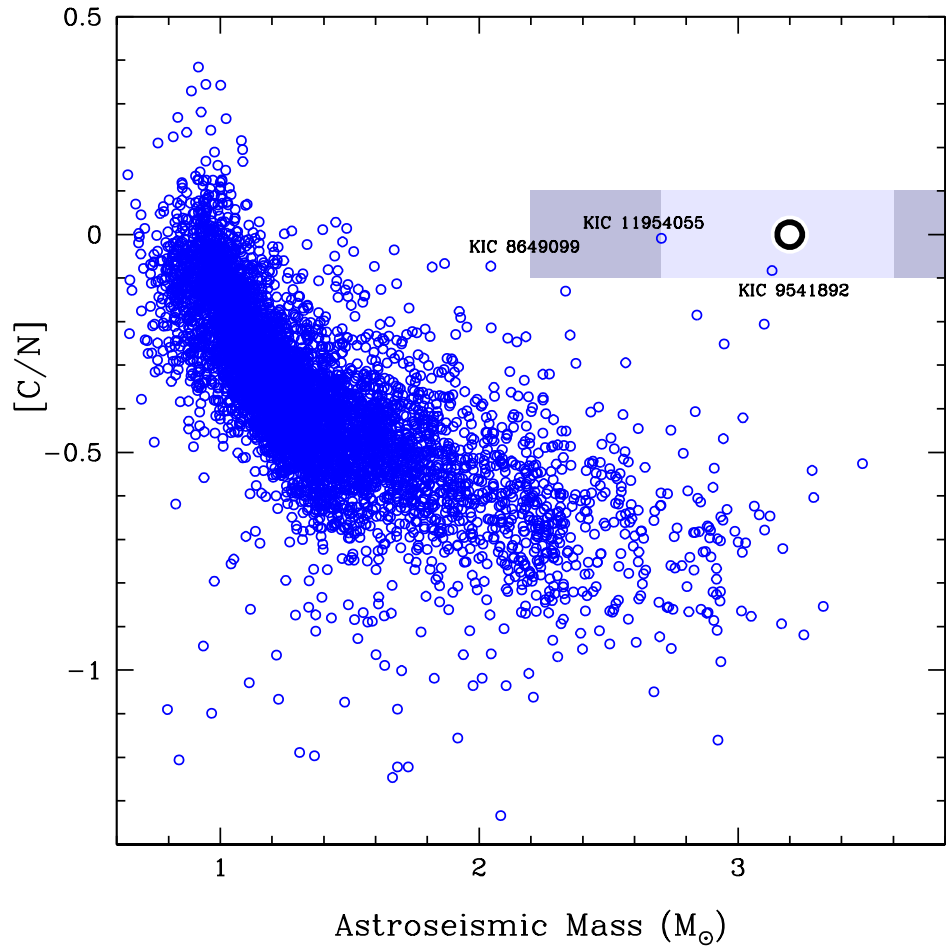}}
\vspace*{-3cm}
\caption{The $[{\rm C/N}]$ ratio from APOGEE as a function of the astroseismic mass in the APOKASC sample \cite{Pinsonneault18}. The $[{\rm C/N}]$ abundance of J05215658 is measured by APOGEE to be $\simeq0.0$ (Table \ref{table:abundances}). The shaded region and the black circle show the mass ranges and best fit for $M_{\rm giant}$ from Figure \ref{figure:m}. Note that the bias in the APOGEE $\log g$ determination (see Section \ref{section:apogee}, Fig.~\ref{figure:rot}) may affect the abundance determinations for J05215658 and other rapidly rotating giants in the APOGEE sample.}
\label{figure:cn}
\end{figure*}

\begin{table}[!t]
\begin{center}
\caption{New and Archival Photometry
\label{table:giant_photometry}}
\begin{tabular}{lccccc}
\hline \hline
\\
\multicolumn{1}{c}{Instrument} &
\multicolumn{1}{c}{Band} &
\multicolumn{1}{c}{Magnitude} &
\multicolumn{1}{c}{Uncertainty} &
\multicolumn{1}{c}{Reference} \\ 
\multicolumn{1}{c}{or Facility} &
\multicolumn{1}{c}{or Filter} &
\multicolumn{1}{c}{or Flux (cgs)} &
\multicolumn{1}{c}{} &
\multicolumn{1}{c}{} \\ 
\\
\hline
\hline
\\
\\
WISE & F34W & 8.73 & 0.05 &  \cite{WISE}\\
          & F46W & 8.79 & 0.05 &   \cite{WISE}\\
          \\

2MASS & Ks & 8.88 & 0.05 &   \cite{2MASS}\\
             & H & 9.07 & 0.05 &   \cite{2MASS} \\
             & J & 9.83 & 0.05 &  \cite{2MASS}  \\
             \\

Post Observatory & $i$  &11.64 & 0.05 &  this work  \\
	        & $r$ & 12.27 & 0.05 &  this work  \\
	        & V & 12.89 & 0.05 & this work   \\
 		& B& 14.34 & 0.05 & this work   \\
		\\

Swift UVOT & U (Vega) & 15.56 &0.04 &this work \\
         & UVM2 (Vega) & 20.00  & 0.095 &  this work \\
         \\

Swift XRT & $0.3-10$\,keV & $<4.4\times10^{-14}$ & &this work\\            
      
 \\
\hline
\hline
\end{tabular}
\end{center}
\end{table}

\section*{Further Acknowledgement}

Funding for the Sloan Digital Sky Survey IV has been provided by the Alfred P. Sloan Foundation, the U.S. Department of Energy Office of Science, and the Participating Institutions. SDSS-IV acknowledges support and resources from the Center for High-Performance Computing at the University of Utah. The SDSS web site is www.sdss.org.

SDSS-IV is managed by the Astrophysical Research Consortium for the  Participating Institutions of the SDSS Collaboration including the  Brazilian Participation Group, the Carnegie Institution for Science,  Carnegie Mellon University, the Chilean Participation Group, the French Participation Group, Harvard-Smithsonian Center for Astrophysics,  Instituto de Astrof\'isica de Canarias, The Johns Hopkins University,  Kavli Institute for the Physics and Mathematics of the Universe (IPMU) / University of Tokyo, Lawrence Berkeley National Laboratory,  Leibniz Institut f\"ur Astrophysik Potsdam (AIP),  Max-Planck-Institut f\"ur Astronomie (MPIA Heidelberg),  Max-Planck-Institut f\"ur Astrophysik (MPA Garching),  Max-Planck-Institut f\"ur Extraterrestrische Physik (MPE),  National Astronomical Observatories of China, New Mexico State University,  New York University, University of Notre Dame,  Observat\'ario Nacional / MCTI, The Ohio State University,  Pennsylvania State University, Shanghai Astronomical Observatory,  United Kingdom Participation Group, Universidad Nacional Aut\'onoma de M\'exico, University of Arizona,  University of Colorado Boulder, University of Oxford, University of Portsmouth,  University of Utah, University of Virginia, University of Washington, University of Wisconsin,  Vanderbilt University, and Yale University.

ASAS-SN is supported by the Gordon and Betty Moore Foundation through grant GBMF5490 to the Ohio State University and NSF grant AST-1515927. Support for ASAS-SN has also come from NSF grant AST-0908816, the Mt.\ Cuba Astronomical Foundation, the Center for Cosmology and Astro-Particle Physics at the Ohio State University, the Chinese Academy of Sciences South America Center for Astronomy (CAS- SACA), the Villum Foundation, and George Skestos. C.S.K. and K.Z.S. are supported in part by NSF grants AST-1515927 and AST-1515876.

This work has made use of data from the European Space Agency (ESA) mission {\it Gaia} (\url{https://www.cosmos.esa.int/gaia}), processed by the {\it Gaia} Data Processing and Analysis Consortium (DPAC, \url{https://www.cosmos.esa.int/web/gaia/dpac/consortium}). Funding for the DPAC has been provided by national institutions, in particular the institutions participating in the {\it Gaia} Multilateral Agreement.

\end{document}